\documentclass{amsart}
\usepackage{amsmath,graphicx}

\usepackage{amsthm}
\usepackage{amssymb}
\usepackage{xypic}
\usepackage{hyperref}
\usepackage[square]{natbib}
\setcitestyle{numbers}
\theoremstyle{plain}

\theoremstyle{definition}

\title[Using an astrolabe to measure orbits]{Astronomy with Chaucer: Using an astrolabe to determine planetary orbits}
\author{Michael Robinson}
\address{Mathematics and Statistics\\
American University\\
Washington, DC, USA}
\email{michaelr@american.edu}

\begin{document}

\begin{abstract}
  Armed with an astrolabe and Kepler's laws one can arrive at accurate estimates of the orbits of planets.
\end{abstract}

\maketitle

\tableofcontents

\section{Introduction}

Without a telescope, even in the midst of the city lights, you can see some of the brightest stars and planets.  Over a span of two or three years, you can watch the planets move across the sky.  Surprisingly, with basic observations of these objects and a simple tool called an \emph{astrolabe} you can tell the time, find compass directions, determine the season, and even determine the diameter of planetary orbits \cite{Morrison_2007,North_1974}.  Astrolabes have been made since antiquity, and you can even make your own.  (That's what I did; see Figure \ref{fig:dissection}.)  My astrolabe follows a design popular in medieval England, whose use is described by the famous poet Geoffrey Chaucer in a letter to his son \cite{Chaucer_astrolabe}.

So if you have an astrolabe, what can you do with it?  How accurate is it?  Can you use an astrolabe to measure the solar system?  Can you verify that the sun is the center of the solar system?  Over the past three years, I have used my astrolabe to collect sightings of celestial bodies visible from my back yard, hoping to answer these questions.  Although the astronomical tool is primitive, the resulting dataset is ripe for modern data processing and yields interesting insights. 

\section{Motivation}

I have had an abiding interest in timekeeping and timekeeping mechanisms.  Mechanical timekeepers have had a storied past, which the reader is encouraged to explore.  Over most of their history, mechanical timekeepers have been rather unreliable at keeping a stable ``rate,'' which is to say that sometimes they run too fast or too slow.  One can reasonably ask, ``too fast or slow with respect to \emph{what}?''  The best source of time until the 20th century had been obtained from the rotation of the Earth, which can be measured by tracking the movement of the stars.  Given that one of the best ways to learn about a technique is to participate in it directly, I set about trying basic astronomical timekeeping techniques.  One thing led to another, and I eventually found myself systematically collecting measurements of various celestial bodies!

Before telescopes became widely available, there were many different instruments to aid in collecting systematic measurements of the sky.  Many of them are challenging to construct accurately.  Construction manuals with accurate translations are difficult to find because of their relative obscurity.  Even if one were to obtain such an instrument, its use may remain impossible because many archaic astronomical instruments were intended to be supplemented with tables that are no longer extant.

The most famous of the astronomical instruments is the \emph{astrolabe}.  It is based upon a stereographic projection of the stars and is self-contained, requiring no additional tables to use.  Stereographic projections are easy to construct, both manually (with a straightedge and compass) or using a computer.  Furthermore, a complete step-by-step guide (addressed to a child, no less!) for using an astrolabe was written circa 1400 AD by Geoffrey Chaucer, \emph{A Treatise on the Astrolabe} \cite{Chaucer_astrolabe}.

Although Chaucer is primarily known now as a master of poetry and fiction, his \emph{Treatise} is apparently the earliest known technical manual written in English \cite{Lipson_1982}!  It is well organized and is written in clear, technical prose.  With a little effort, it can be understood by a modern reader in its original form, especially if one refers to an actual astrolabe.

Although astrolabes can be easily fabricated using a computer using a variety of materials and can collect usefully accurate data about the sky, they are not widely \emph{used} to collect and analyze these data.  Instead, the once-noble astrolabe appears to feature only in historical discussions in the classroom \cite{etal2021AstrolabeAL,de2015workshop,Ford_2012,Zotti2008TangibleHP,Winterburn_2005,Ransom_1993} (though \cite{wlodarczyk_1987} appears to be a notable exception).  It was originally for this reason---to understand the \emph{Treatise} and its relation to timekeeping \emph{through daily usage}---that I undertook to fabricate an astrolabe.

\section{Design of the astrolabe}
\label{sec:design}

\begin{figure}
  \begin{center}
    \includegraphics[width=3in]{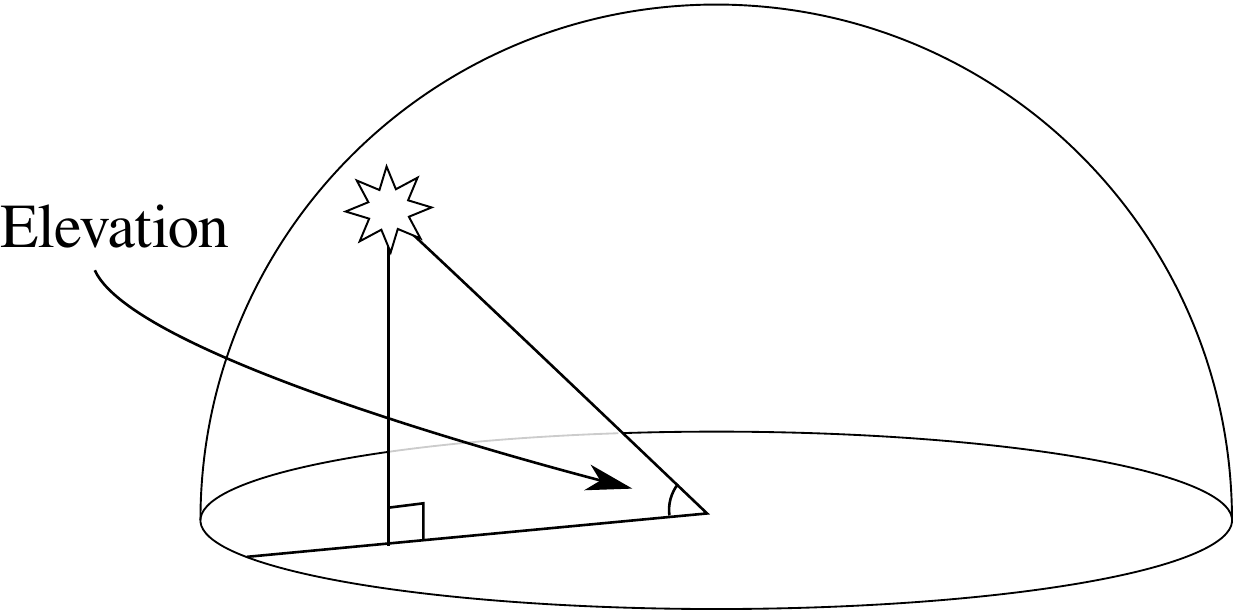}
    \caption{Taking elevation sightings of a object in the sky.}
    \label{fig:sightings}
  \end{center}
\end{figure}

The main use of an astrolabe is to measure the elevation angle above the horizon of celestial bodies (Figure \ref{fig:sightings}), and from these \emph{elevation sightings} derive other quantities of interest.  The astrolabe described by Chaucer is a rather complete instrument, with an astonishing variety of capabilities.  Not all of these are necessary to determine planetary orbits.  For that task, one only needs a set of observations containing
\begin{enumerate}
\item the \emph{local time} and date, and
\item the \emph{right ascension} of each visible planet.
\end{enumerate}
Both of these quantities can be derived from elevation sightings, using the astrolabe as a mechanical calculator as explained in Section \ref{sec:predicting_true_locations}.
With these two quantities in hand, Kepler's third law can be used to determine the semi-major axis (roughly speaking, the radius) of each planet's orbit as will be explained in Section \ref{sec:orbits}.

\subsection{Right ascension and declination}
\label{sec:celestial_sphere}

\begin{figure}
  \begin{center}
    \includegraphics[width=4in]{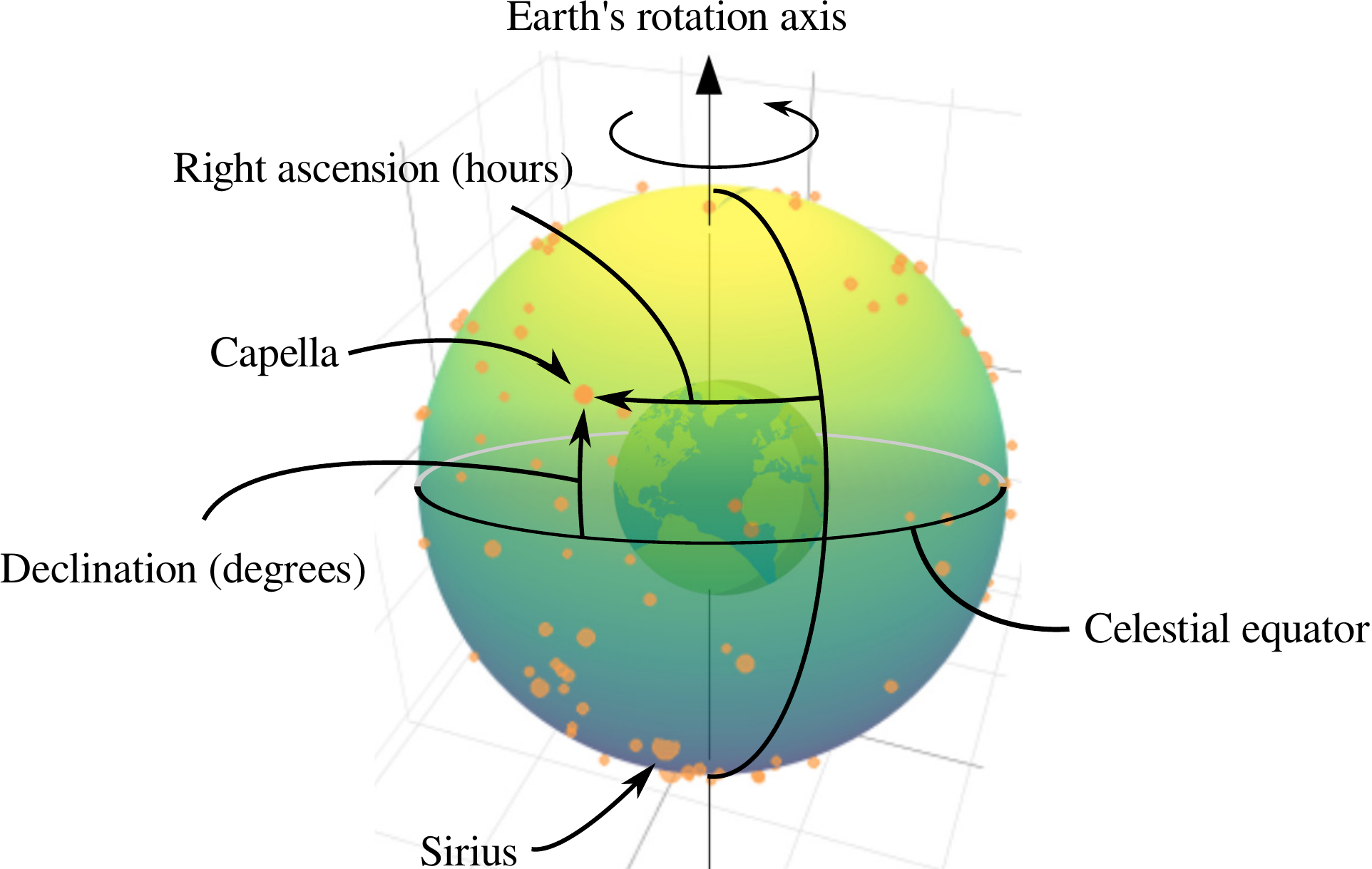}
    \caption{The celestial sphere, showing right ascension and declination coordinates of the brightest stars.  The star Cappella has right ascension 5 hours 17 minutes and declination $45.83^\circ$.  The star Sirius has right ascension 6 hours 45 minutes and declination $-16.72^\circ$.} 
    \label{fig:celestial_sphere}
  \end{center}
\end{figure}

Since the stars are very far from the Earth, they appear as points of light that do not move over the course of the year.
Because of this, we can imagine that each star is projected onto the surface of an immobile sphere (Figure \ref{fig:celestial_sphere}), called the \emph{celestial sphere}.
We can therefore specify the location of each star in the sky using two angular coordinates on the celestial sphere, called \emph{right ascension} and \emph{declination}.
Declination on the celestial sphere is akin to latitude on the Earth, with the Earth's true North pole being given a declination of $90^\circ$.
Stars directly over Earth's equator are given a declination of $0^\circ$, and so on.
Right ascension on the celestial sphere is akin to longitude on the Earth, though it is typically measured in \emph{hours} instead of degrees.
There are $15^\circ$ per hour of right ascension.
By convention, zero right ascension is in the direction specified by the intersection of the Earth's equatorial plane and its orbital plane around the Sun, pointing away from the Sun through the Spring equinox.  This rather complicated definition does not concern us here, aside from the fact that there \emph{is} a well-defined convention!

\subsection{Stereographic projection}
\label{sec:stereographic}

As the Earth rotates on its axis,
celestial bodies appear to rise in the East and set in the West each day.
Over the course of the year, the Earth orbits the Sun at a roughly constant speed along a nearly perfect circle.
These two time scales---the daily and yearly cycles---make it complicated to track the motion of celestial bodies.
If one holds the Earth's rotation angle fixed as it moves along its orbit, the stars will appear to remain fixed.
The motion of the Sun and the planets against this fixed background of stars becomes easier to study.
The astrolabe is a mechanical calculator that enables one to measure positions of celestial bodies against this background of fixed stars.

\begin{figure}
  \begin{center}
    \includegraphics[width=5in]{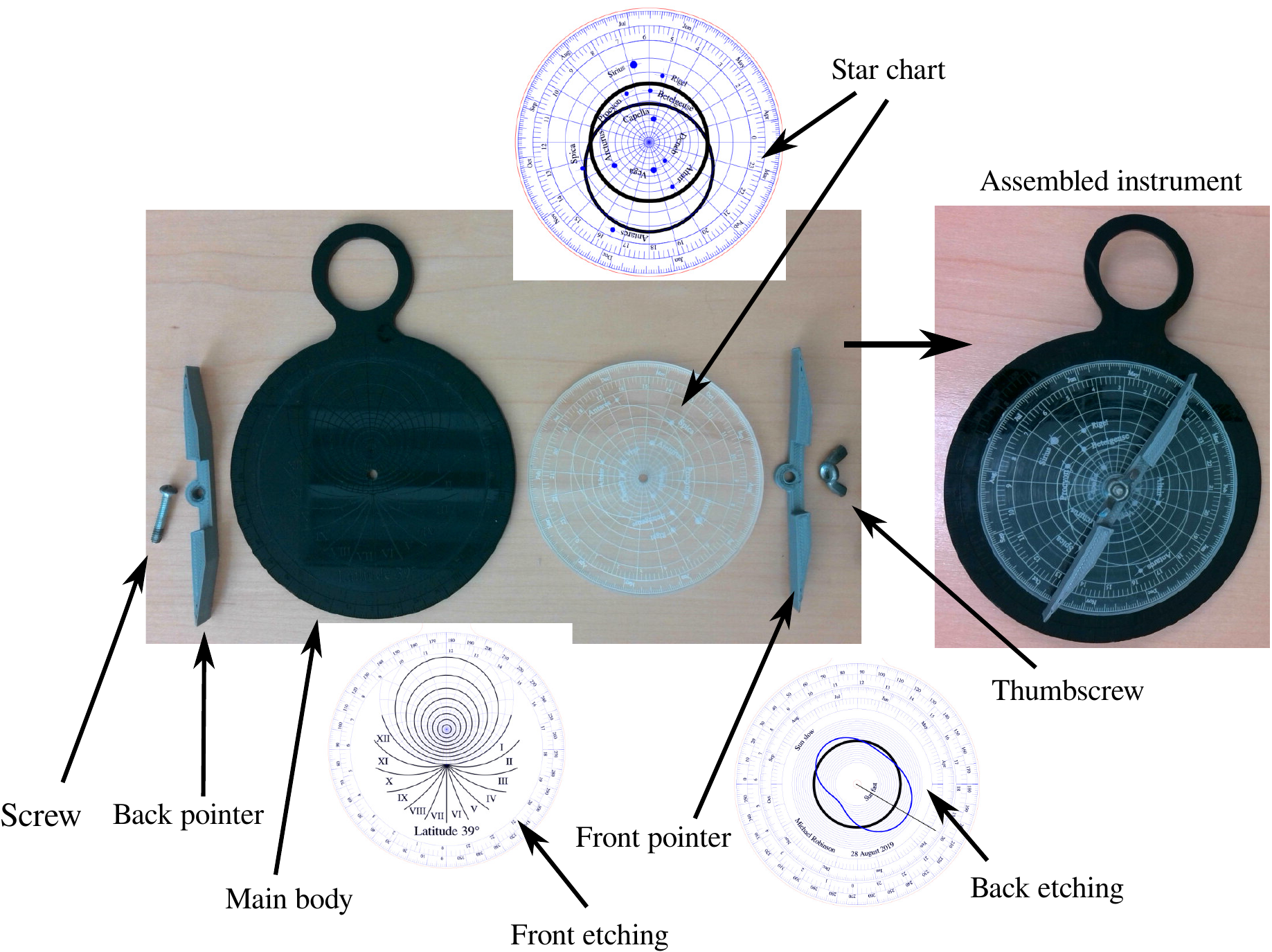}
    \caption{One of my astrolabes (the {\tt small\_astrolabe} \cite{Robinson_small_astrolabe}) used in collecting my dataset.}
    \label{fig:dissection}
  \end{center}
\end{figure}

The front of my astrolabe consists of two pieces, pinned together with a screw as shown in Figure \ref{fig:dissection}.
The main body of the instrument is usually held stationary and contains etchings of elevation contours.
The stars are etched on a clear sheet of plastic---the \emph{star chart}---which shows the locations of some easy-to-find stars in the background.
The star chart is free to rotate.
As the Earth rotates over the course of a day, the star chart makes slightly more than one full rotation about the main body.

\begin{figure}
  \begin{center}
    \includegraphics[width=5in]{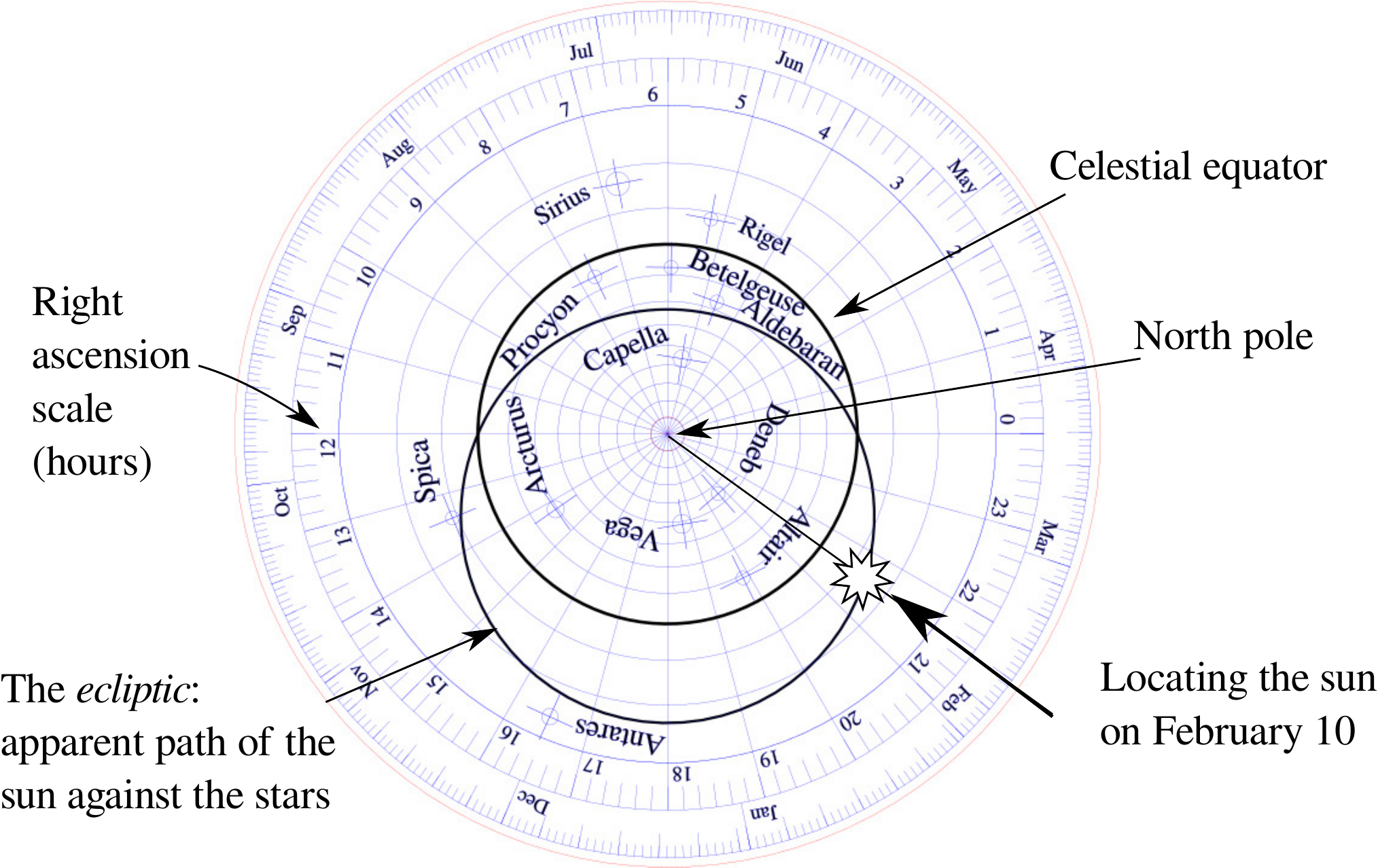}
    \caption{The star chart of the astrolabe, showing the guide stars and the method for locating the Sun.}
    \label{fig:star_chart}
  \end{center}
\end{figure}

The right ascension and declination of each star specifies its coordinates on the star chart via polar coordinates and \emph{stereographic projection}.
The angular coordinate is specified by the right ascension, which is indicated around the circumference of the star chart (Figure \ref{fig:star_chart}).
Since I live in the northern hemisphere, the north pole at declination $90^\circ$ is at the center screw.
The celestial equator is a circle that is concentric with the screw.
Stars with positive declinations (for instance, Capella) appear inside the celestial equator on the star chart,
while stars with negative declinations (for instance, Sirius) appear outside the celestial equator.
Because the star chart is drawn with stereographic projection, angles (but not areas) are preserved.
This makes it possible to identify certain constellations on the astrolabe because their stars are in their correct spatial relation to each other.
Additionally, stereographic projection means that circles and lines drawn on the star chart correspond to lines or circles in the sky.  

One can also make an astrolabe for the southern hemisphere.  In that case, one would set the south pole (declination $-90^\circ$) at the screw,
and the stars would appear in different locations on the star chart as a result.

\subsection{The ecliptic}
\label{sec:ecliptic}

Holding the stars fixed, Earth's motion along its orbit means that the Sun appears to move through the sky.
Aside from the daily sunrise in the East and sunset in the West,
the Sun also appears to move against the background of immobile stars in the opposite direction: from West to East.
In actuality, one cannot see the stars that appear near the Sun (except during a solar eclipse), but its motion can be inferred from the stars that are visible.
Over the course of one year, the Sun returns to its original location against the background of the stars.

Since the star chart is drawn with stereographic projection,
any plane in space that is concentric with the Earth becomes a circle or a line on the star chart.
The Earth's orbit lies in a plane containing both the Earth and the Sun
and is inclined roughly $22.5^\circ$ with respect to the celestial equator.
Because of this, the path that the Sun appears to follow through the sky is a circle on the star chart of the astrolabe, called the \emph{ecliptic}.
The inclination results in an offset of the center of the ecliptic from the North pole.

The Sun's right ascension increases with a mean rate of approximately $(1/365.25)^\circ$ per day.
The star chart shown in Figure \ref{fig:star_chart} has a calendar scale plotted around its circumference that allows the user of the astrolabe to place the Sun at its proper location along the ecliptic for any given date.  This is explained in further detail in Section \ref{sec:placing_sun}.

The instantaneous rate of motion of the Sun varies from this mean rate over the course of the year,
resulting in a maximum difference of roughly $16$ minutes of right ascension on some dates.
Section \ref{sec:equation_of_time} describes a formula called the \emph{equation of time} that models these variations.
I included these variations on the star chart, hoping that they might improve the accuracy of my time estimates.
As I later found (see Section \ref{sec:ra_error}), the difference between the modern formula and the one Chaucer used is much smaller than what can be reliably measured using the astrolabe.

\subsection{Guide stars}
\label{sec:guide_stars}

I live in a light-polluted suburb of Washington, DC, from which only a few bright stars are visible\footnote{My observing location tends to be either 8 or 9 on the Bortle scale \cite{Bortle}, where 1 is the darkest sky and 9 is the most light-polluted sky.}. 
Nevertheless, along with the planets Venus, Mars, Jupiter, and Saturn, enough stars are visible to derive useful information.
After a few sessions of observing, I determined that $12$ frequently visible stars could serve as my ``guide stars.''
These stars are the ones that I plotted on the star chart of my astrolabe (Figure \ref{fig:star_chart}).
I endeavored to record elevation sightings for each of these guide stars whenever possible.  This forms the bulk of my data.

Determining star positions in right ascension and declination has been an important task for astronomers since antiquity.
All $12$ of my guide stars have published right ascension and declination coordinates from the European Space Agency's Hipparcos mission.
I used a somewhat simplified version of these data, dervied from the Hippacos v3 stellar catalog \cite{hyg3scv}.
If these high-quality data were not available, one could resort to using the astrolabe to make these measurements, since Chaucer gives detailed instructions for this task \cite[II.17-18]{Chaucer_astrolabe}.

\subsection{Fabrication considerations}
\label{sec:fabrication}

I designed and built my astrolabe (Figure \ref{fig:dissection}) according to the description in Part I of the \emph{Treatise},
since Chaucer gives a very complete description of his instrument.
I made a few changes to Chaucer's plan to suit my preference for modern timekeeping.
Most notably, having ready access to other computational tools,
I did not need the compact trigonometric table Chaucer includes on the back of his instrument.
In its place, I included the modern \emph{equation of time},
which appears as the peanut-shaped curve on the ``Back etching'' in Figure \ref{fig:dissection} and is described in Section \ref{sec:equation_of_time}.
The equation of time enables estimation of longitude as described in Section \ref{sec:longitude}.

The reader is encouraged to construct their own astrolabe.
There are a variety of sources of plans suitable for printing on paper or laser cutting in plastic, for instance see \cite{Astrolabe_project,Zotti2008TangibleHP}.

The plans for the instrument I used to collect the data in this paper are available on GitHub:
\begin{itemize}
\item the version that I used to collect most of the data in this article: \cite{Robinson_large_astrolabe}, and
\item a ``pocket size'' portable version \cite{Robinson_small_astrolabe} (shown in Figure \ref{fig:dissection}).
\end{itemize}
There are several ``branches'' in the GitHub repositories that contain designs for astrolabes for certain specific latitudes.
I have constructed and used astrolabes according to these plans using various materials, including laminated paper, thin plywood, and cast acrylic plastic.
My first astrolabe consisted of printed sheets of paper glued to thin plywood and could be made without access to a sophisticated workshop.
The dataset described in this article uses both sizes of plastic astrolabes listed above, with no noticeable difference in accuracy.

Although the GitHub repositories listed above contain files that are ready for fabrication, modifications can be easily made.
All the graphics on both astrolabes were generated by a Python Jupyter notebook that is included with the plans.
Therefore, if you wish to make an astrolabe for a different latitude or apply some other change to the instrument,
you can modify and rerun the corresponding notebook.
The Jupyter notebook handles all the calculations, graphics, and most of the labeling.
For purely aesthetic reasons, I extensively retouched the graphics files the notebook produces.
The most difficult part of the retouching process is aligning the thumb ring,
which acts as a key to align the front and back of the instrument while it is being constructed.

\section{Using the astrolabe}
\label{sec:predicting_true_locations}

\begin{figure}
  \begin{center}
    \includegraphics[width=6in]{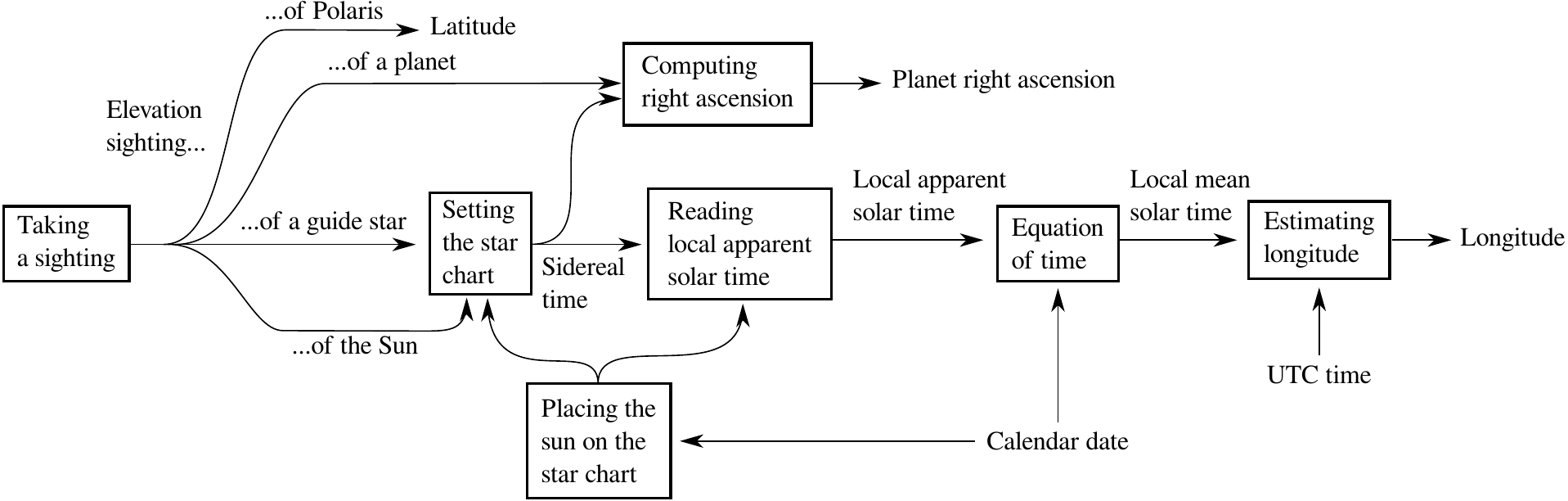}
    \caption{Flow chart of astrolabe tasks and measurements.  Boxed items indicate specific tasks to be performed with the astrolabe as described in the text.  Unboxed items indicate measurements.}
    \label{fig:flowchart}
  \end{center}
\end{figure}

The astrolabe is a capable instrument, but it is not completely straightforward to use.
While the reader is encouraged to read Chaucer's \emph{Treatise} in its entirety, this paper outlines the basic tasks necessary for measuring the orbits of the planets.
Some of the tasks that one needs to accomplish depend on others, as is shown in Figure \ref{fig:flowchart}.  These are discussed in more detail in the following sections:
\begin{itemize}
\item placing the sun on the star chart (Section \ref{sec:placing_sun}),
\item taking a sighting (Section \ref{sec:taking}),
\item setting the star chart (Section \ref{sec:setting}),
\item reading local apparent solar time (Section \ref{sec:reading}),
\item equation of time (Section \ref{sec:equation_of_time}),
\item estimating longitude (Section \ref{sec:longitude}), and
\item computing right ascension (Section \ref{sec:computing_ra}).
\end{itemize}

\subsection{Kinds of time}
\label{sec:time}

My initial motivation for studying astrolabes was timekeeping.
The astrolabe serves admirably in this role.
Given the wide availability of an accurate global time, Universal time (UTC), it is useful to recall how our timekeeping is tied to the motion of celestial bodies.

Consider a sundial, an instrument that measures time with a shadow cast by the Sun.
The shadow cast on a sundial appears to move in time with the Sun's position in the sky.
Sundials measure \emph{local apparent solar time}.
The length of an \emph{apparent solar day} is the time interval over which the shadow returns to a given location on the sundial.  

The length of the apparent solar day varies over the course of the year, though not more than about 30 seconds.
This variation means that the Sun's apparent motion through the sky is not quite at a constant rate.

Variation in the length of the apparent solar day means that clocks---which do run at a constant rate---do not measure apparent solar time.
Instead, clocks are set using the \emph{mean} length of the apparent solar day over the course of the year.
The corresponding averaged day is called the \emph{mean solar day}, and results in \emph{mean solar time}.
The mean solar time and apparent solar time differ by an amount that is the accumulation of the variations in apparent solar day lengths over many days (or months), and can amount to twenty minutes or so at various points during the year.
The \emph{equation of time}, described in Section \ref{sec:equation_of_time}, models this difference.

The astrolabe directly displays the apparent solar time \cite[II.3]{Chaucer_astrolabe}, as described in Section \ref{sec:reading}.
From the the apparent solar time, it is possible to determine mean solar time by using the equation of time.

Finally, since the Earth is roughly spherical, the apparent position of the Sun in the sky at any given time depends on one's longitude.
The times discussed above are therefore \emph{local} to one's longitude.
By international agreement, UTC is\footnote{To within the accuracy of my astrolabe, anyway!} the local mean solar time at $0^\circ$ longitude.

\subsection{Placing the sun on the star chart}
\label{sec:placing_sun}


One of the most basic tasks of the astrolabe is to predict the position of the Sun along the ecliptic.
On the front face of the astrolabe, the ecliptic is an offset circle, along which the Sun moves counterclockwise at a nearly constant rate,
making one full circuit each year.
Chaucer's astrolabe divided this circle into 365 equal intervals, one for each day.  
To locate the Sun on a given date, one traces a ray starting at the North pole, out to the date shown on the outer ring (see Figure \ref{fig:star_chart}).
Where this line intersects the ecliptic is the Sun's location \cite[II.1]{Chaucer_astrolabe}.

\subsection{Placing the planets on the star chart}
\label{sec:placing_planets}

Positioning the planets on the astrolabe is much harder than positioning the Sun.
Unlike the Sun, the apparent path of a planet is not uniform motion along the ecliptic.
Fortunately, all of the planets' orbits are confined to a few degrees of the ecliptic (see Table \ref{tab:planet_inclinations}).
Chaucer planned to explain how to do this using geocentric orbit theory (presumably relying upon Ptolemy's model) in Part IV of his \emph{Treatise},
though he never got around to writing beyond Part II.
We will never know exactly what his instructions would have been.

Given some preliminary measurements, it is possible to measure the right ascensions of the planets (in Section \ref{sec:computing_ra}),
which places them on the ecliptic.
Nevertheless, according to Kepler's laws (which were discovered after Chaucer),
predicting the locations of the planets is not necessary to find the dimensions of their orbits.
However, once the dimensions have been determined, we can estimate the locations of the planets, as will be shown in Section \ref{sec:orrery}.

\subsection{Taking a sighting}
\label{sec:taking}

The astrolabe incorporates an elevation sighting scale on its back.
Taking a sighting with an astrolabe is basically the same as taking one using a theodolite, sextant, or similar instrument.
One sights a desired star along the attached back pointer, and then reads the corresponding scale \cite[II.2]{Chaucer_astrolabe}.
Amusingly (to me, anyway), Chaucer concludes his brief instructions on how to take a sighting by saying \emph{``This chapitre is so general ever in oon, that ther nedith no more declaracioun; but forget it nat.''}
Loosely translated, ``We'll be using this so frequently that I won't bother to mention taking elevation sightings again, but don't forget how to do it.''
This may be true, but taking accurate elevation sightings required more practice than I expected!

Sighting the Sun requires a special note of caution,
since one should not look directly at the Sun.
The correct procedure is to cast a shadow of the astrolabe.
One then minimizes the apparent size of the shadow of the elevation pointer,
and from this reads the elevation of the Sun.

Even if one does not learn anything else about the astrolabe,
taking elevation sightings is a useful navigational skill since it can be used to determine one's latitude.
To the accuracy of my astrolabe (which is about $2^\circ-3^\circ$, according to Table \ref{tab:elevation_error} in Section \ref{sec:elevation_error}),
one merely needs to sight the elevation of the star Polaris, which is usually called ``the North star''.

Interestingly, Polaris was about $4^\circ$ away from the North pole during Chaucer's time,
yielding an angular difference that is detectible with my astrolabe.
Chaucer describes three options \cite[II.23-25]{Chaucer_astrolabe},
each of which is a rather more elaborate process than sighting Polaris.
While these options are useful when locating the North pole using a telescope,
we do not need to discuss them here.

\subsection{Setting the star chart}
\label{sec:setting}

\begin{figure}
  \begin{center}
    \includegraphics[width=3.5in]{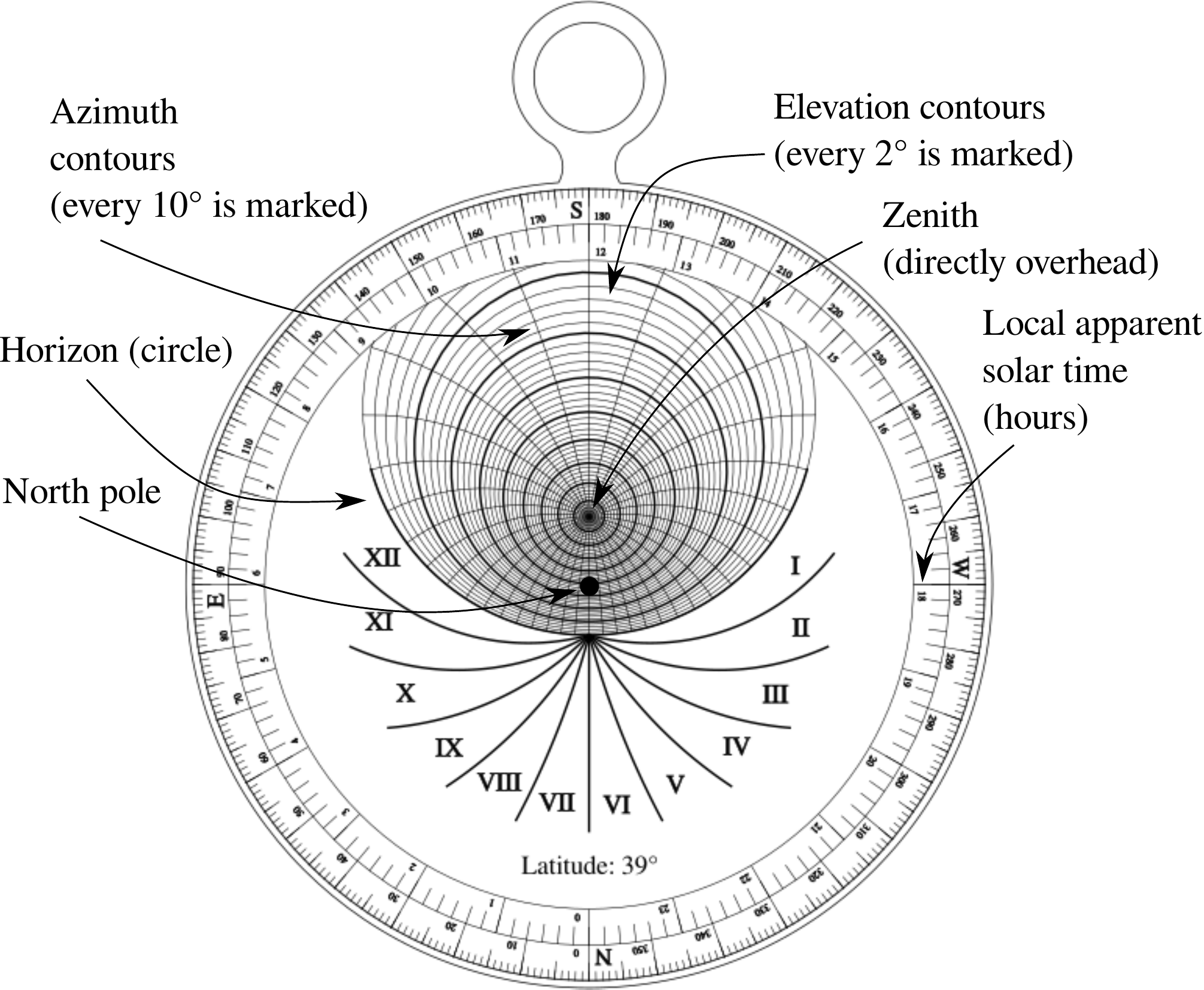}
    \caption{Front markings on the main body of the astrolabe}
    \label{fig:front}
  \end{center}
\end{figure}

The star chart and main body of the astrolabe are pinned together at the North pole (see Figure \ref{fig:dissection}).
The star chart can be rotated around this axis against the main body, yielding a single degree of freedom.
The star chart makes one full rotation in slightly less than 24 hours, in time with the rotation of the Earth.
This enables the Sun or visible stars to be located at the correct elevation contours (see Figure \ref{fig:front}).
Once this has been done correctly, \emph{all} of the stars (and the ecliptic) will be placed at their proper location in the sky.

To perform this task, one begins with an elevation sighting of a known star (or the Sun).
Locate the star on the star chart.
If the sighting was taken of the Sun, find its position on the star chart as described in Section \ref{sec:placing_sun}.
Locate the elevation contour circle on the main body of the astrolabe (see Figure \ref{fig:front}).  It helps to count inwards from the horizon circle, since the horizon is at elevation $0^\circ$.
With these two data in mind, rotate the rotate star chart so that the star is placed on its correct elevation contour.

Since two circles usually intersect in two places, there is an ambiguity.
The star could be located at two different places on the same elevation contour.
Resolve the ambiguity by repeating the process for an elevation sighting of a second object,
or by determining whether the star is closer to the East or the West in the sky.

\subsection{Reading the local apparent solar time}
\label{sec:reading}


Once the star chart has been properly set, the local apparent solar time can then be read from the astrolabe.
Trace a ray from the North pole through the location of the Sun (even if it is not visible) out to the local apparent solar time scale shown on Figure \ref{fig:front}.  Where the ray intersects the local apparent solar time scale is the local apparent solar time corresponding to the position of the sun and stars in the sky.  

\subsection{Equation of time}
\label{sec:equation_of_time}

Local mean solar time can be obtained from the local apparent solar time by using the \emph{equation of time}, which is the offset between local apparent solar time and local mean solar time \cite{eqtime}.  For the purposes of this article, the offset is approximately 
\begin{equation*}
  \Delta t = -7.659 \sin M  + 9.863 \sin (2M + 3.5932) \text{ minutes}
\end{equation*}
where
\begin{equation*}
  M=6.24004077 + 0.01720197 D \text{ radians},
\end{equation*}
and $D$ is the number of days since 1 January 2000.
Therefore, the local mean time is simply obtained by adding $\Delta t$ to the local apparent solar time obtained in Section \ref{sec:reading}.

Notice that if $D=365.24$ days, then $0.01720197 D \approx 2\pi$.  To a good approximation, $\Delta t$ is periodic with a one year period.  The back face of my astrolabe (see Figure \ref{fig:dissection}) has a curve (drawn in polar coordinates) that approximates $\Delta t$ based on the date.  To use this curve, one traces a ray from the center of the instrument to the desired date.  The offset $\Delta t$ can be read where this ray intersects the curve.


\subsection{Estimating longitude}
\label{sec:longitude}

It was long known that a difference in longitude results in a difference in local apparent solar time (or local mean solar time) \cite[II.39]{Chaucer_astrolabe}.  
Chaucer does not explain how to calculate longitudinal difference between two locations,
because it requires a shared time reference between both locations.
That would not become available until 200 years later with Galileo's discovery of the eclipses of Jupiter's moons (January 1610) \cite{sobel2005longitude}.

UTC corresponds to local mean time at $0^\circ$ longitude and is now readily available.  Longitude is therefore obtained by subtracting the local mean time determined from the UTC time, and converting to degrees of longitude by multiplying by $15^\circ$ per hour.

\subsection{Computing planetary right ascensions}
\label{sec:computing_ra}
  
Once the star chart is correctly rotated against the main body of the astrolabe, the right ascensions of the planets can also be determined.
This relies on a special feature of our solar system.
All the planets have orbital planes that are quite close to one another, to within $5^\circ$ (see Table \ref{tab:planet_inclinations}).
To observational accuracy, the planets will appear close to (or on) the ecliptic on the star chart of the astrolabe.
(The Earth's Moon also lies within $5^\circ$ of the ecliptic.)
Therefore, to determine the right ascension of a planet, Chaucer instructs one to intersect the ecliptic with the corresponding elevation contour on the front of the astrolabe \cite[II.34]{Chaucer_astrolabe}.

\begin{table}
  \begin{center}
        \caption{Orbital plane inclination compared to the ecliptic \cite{trefil2012space}}
    \label{tab:planet_inclinations}
    \begin{tabular}{|l|l|l|l|l|l|}
      \hline
      Planet&Venus&Earth&Mars&Jupiter&Saturn\\
      \hline
      \hline
      Inclination&$3.4^\circ$&$0^\circ$ (by def'n)&$1.9^\circ$&$1.3^\circ$&$2.5^\circ$\\
      \hline
    \end{tabular}
  \end{center}
\end{table}

Since the ecliptic and the elevation circles intersect twice, there is ambiguity.
Both intersections may be plausible.
Chaucer warns the reader to avoid this ambiguity by paying attention to which side of the meridian the planet lies on.
As will be later shown in Section \ref{sec:ra_error}, I was not always careful!

\section{Description of my dataset}

\begin{figure}
  \begin{center}
    \includegraphics[width=3.5in]{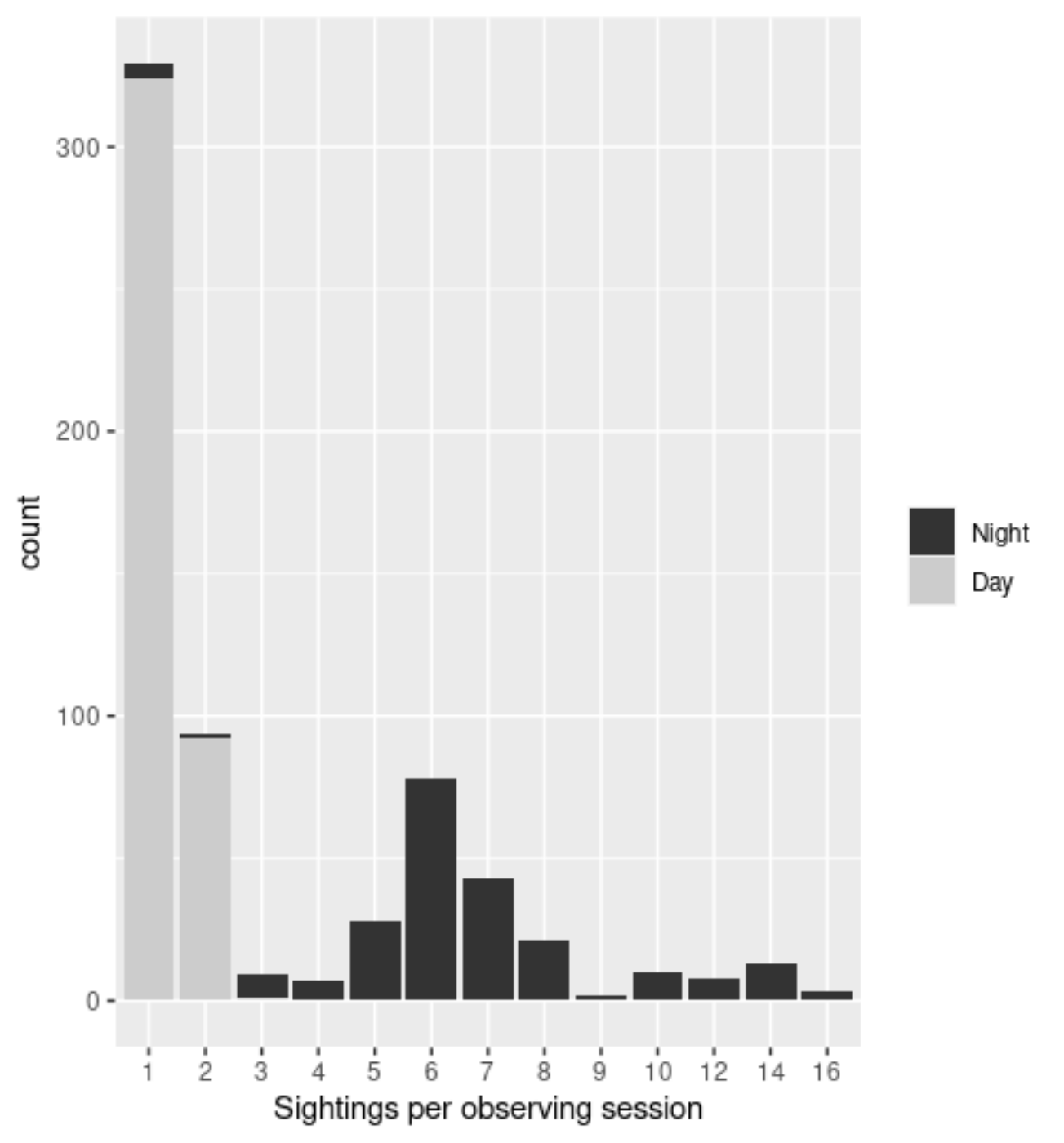}
    \caption{Number of elevation sightings for each of the $773$ observing session.}
    \label{fig:dataset_summary}
  \end{center}
\end{figure}

Using astrolabes I fabricated, I made $773$ observing sessions over the course of three years.
Although the number of visible objects varied substantially, I made $2093$ distinct elevation sightings as summarized in Figure \ref{fig:dataset_summary} and Table \ref{tab:dataset_summary}.
In each session, I recorded the current time from my computer's clock,
the elevations of all visible planets,
and the elevations of all visible guide stars.
The reader is encouraged to examine the data and the associated analysis script,
which are freely available \cite{Robinson_astrolabe_analysis}.

With the exception of the daytime sightings of Venus, I used no optical aid other than the astrolabe.
Perhaps surprisingly, Venus can be seen during the daytime without optical aid under ideal conditions.
However, this requires knowing where to look first!  It is much easier to locate Venus with binoculars first.
Once sighted with the binoculars, the elevation angle to Venus can then be sighted (without binoculars) using the astrolabe.
Since this is a somewhat challenging procedure, I have documented it in a video \cite{Robinson_venus}.

\begin{table}
    \begin{center}
    \caption{Number of elevation sightings collected by object}
    \label{tab:dataset_summary}
    \begin{tabular}{|l|c|r|r|}
      \hline
      Object&Type&Day&Night\\
      \hline
      \hline
      Sun&&494&\\
      \hline
      Venus&Planet&15&32\\
      \hline
      Mars&Planet&&72\\
      \hline
      Jupiter&Planet&1&110\\
      \hline
      Saturn&Planet&1&89\\
      \hline
      \hline
      Aldebaran&Star&&67\\
      \hline
      Altair&Star&&126\\
      \hline
      Antares&Star&&52\\
      \hline
      Arcturus&Star&&134\\
      \hline
      Betelgeuse&Star&&98\\
      \hline
      Capella&Star&&139\\
      \hline
      Deneb&Star&&149\\
      \hline
      Procyon&Star&&96\\
      \hline
      Rigel&Star&&80\\
      \hline
      Sirius&Star&&75\\
      \hline
      Spica&Star&&88\\
      \hline
      Vega&Star&&175\\
      \hline
      \hline
      Total&& 511& 1582\\
      \hline
    \end{tabular}
    \end{center}
\end{table}

\section{Measuring the accuracy of the astrolabe}

Before embarking upon the task of computing the orbits of the planets, it is useful to determine the accuracy of the measurements one can collect with my astrolabe.
Since the process for deriving time and right ascension measurements involves several steps, it is important to check the accuracy at several points along the process shown in Figure \ref{fig:flowchart}.

\subsection{Elevation errors}
\label{sec:elevation_error}

\begin{figure}
  \begin{center}
    \includegraphics[width=5in]{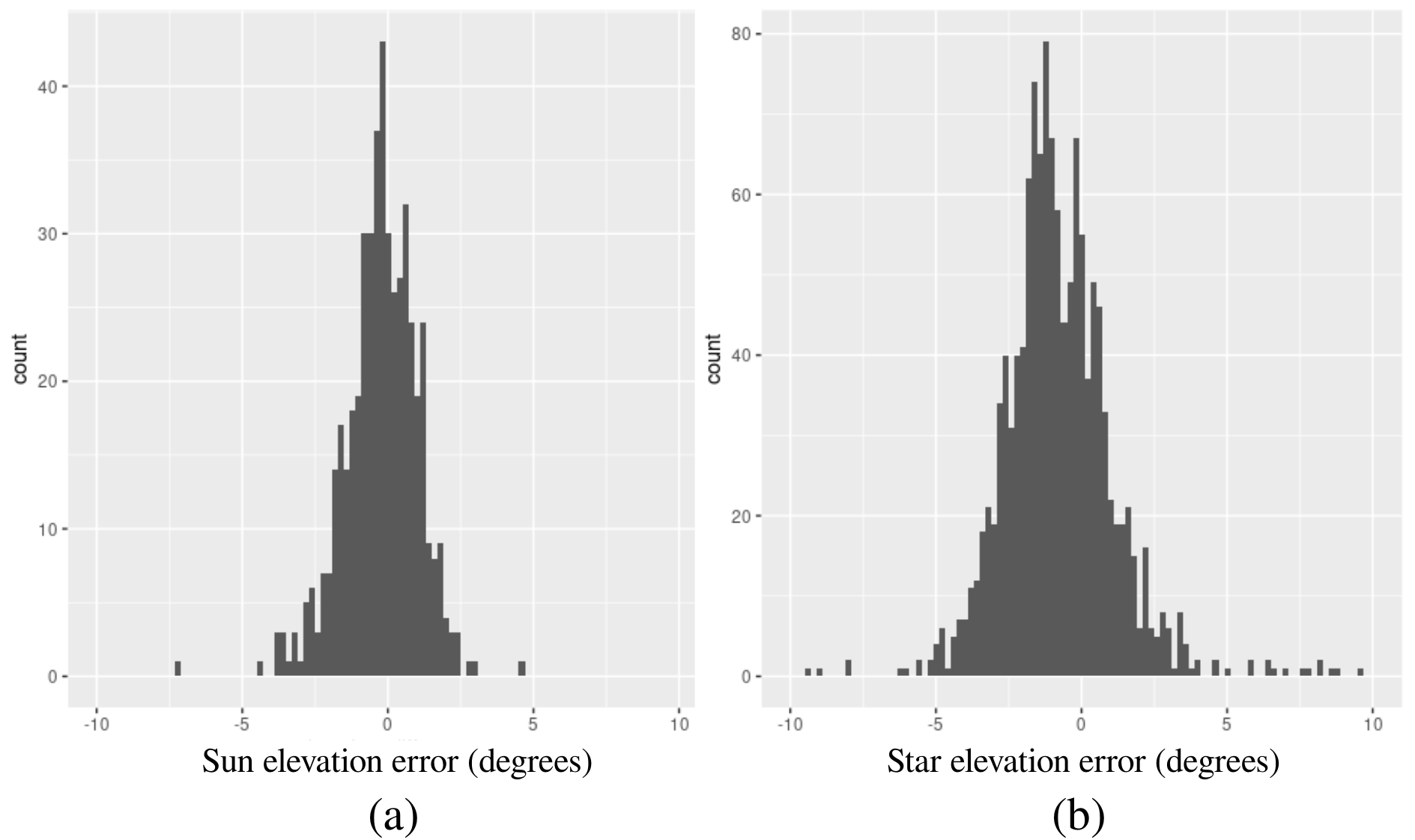}
    \caption{Histogram of elevation errors (astrolabe observed elevation minus expected elevation based on time) using (a) the Sun and (b) the stars.}
    \label{fig:elevation_error}
  \end{center}
\end{figure}

Using the positions of the Sun and and stars as described in Section \ref{sec:predicting_true_locations}, it is an exercise in spherical trigonometry to translate these positions into elevation angles at any given location and time on the Earth.  These values can be compared with the elevation sightings collected with the astrolabe.  The overall statistics are summarized in Table \ref{tab:elevation_error}, and histograms for the Sun and stars are displayed in Figure \ref{fig:elevation_error}.  The distributions of errors for the individual stars and planets are quite similar to each other, and are therefore not shown.  The histograms suggest that the elevation errors follow a normal distribution, which can be taken as an indication that the errors are not biased.

\begin{table}
  \begin{center}
    \caption{Errors in elevation sightings (degrees)}
    \label{tab:elevation_error}
    \begin{tabular}{|l|r|r|r|}
    \hline
    Observation & $n$ & Mean & Std. dev.\\
    \hline
    \hline
    Sun (day) & $494$ & $-0.37^\circ$ & $2.6^\circ$\\
    \hline
    Stars (night) & $1279$ & $-0.75^\circ$ & $2.6^\circ$\\
    \hline
    \end{tabular}
  \end{center}
\end{table}

A two-sample $t$-test from Table \ref{tab:elevation_error} yields $p=0.005$, so there is a significant difference in mean elevation estimation error for the Sun versus the stars.  In short, the Sun estimates are a little more centered than those of the stars.  I suspect that this difference has to do with the design of my particular instrument and the techniques I used.

To sight a star, the user must ``skim'' the star along the top edge of the pointer.  This tends to bias the estimates lower in elevation because an overestimate means that the pointer obstructs view of the star.  On the other hand, one cannot look at the Sun directly.  As explained in Section \ref{sec:taking}, to measure the Sun, one casts a shadow with the pointer and tries to minimize its apparent size.  The procedure for measuring the Sun appears to have less elevation bias than directly sighting a star.

Interestingly, a related instrument, \emph{the mariner's astrolabe}, is optimized for sighting the Sun.
The pointer on the mariner's astrolabe has two vanes, each perforated a tiny hole through which the sunlight may pass only when the correct elevation is displayed.
The mariner's astrolabe can be used to obtain substantially better accuracy than what I was able to collect, with accuracies of $0.1^\circ$ being typical for skilled users on land \cite{Hilster_2014, Koeberer_2014, Knox_johnston_2013}.   

\subsection{Time estimation accuracy}
\label{sec:mean_time}

\begin{figure}
  \begin{center}
    \includegraphics[width=5in]{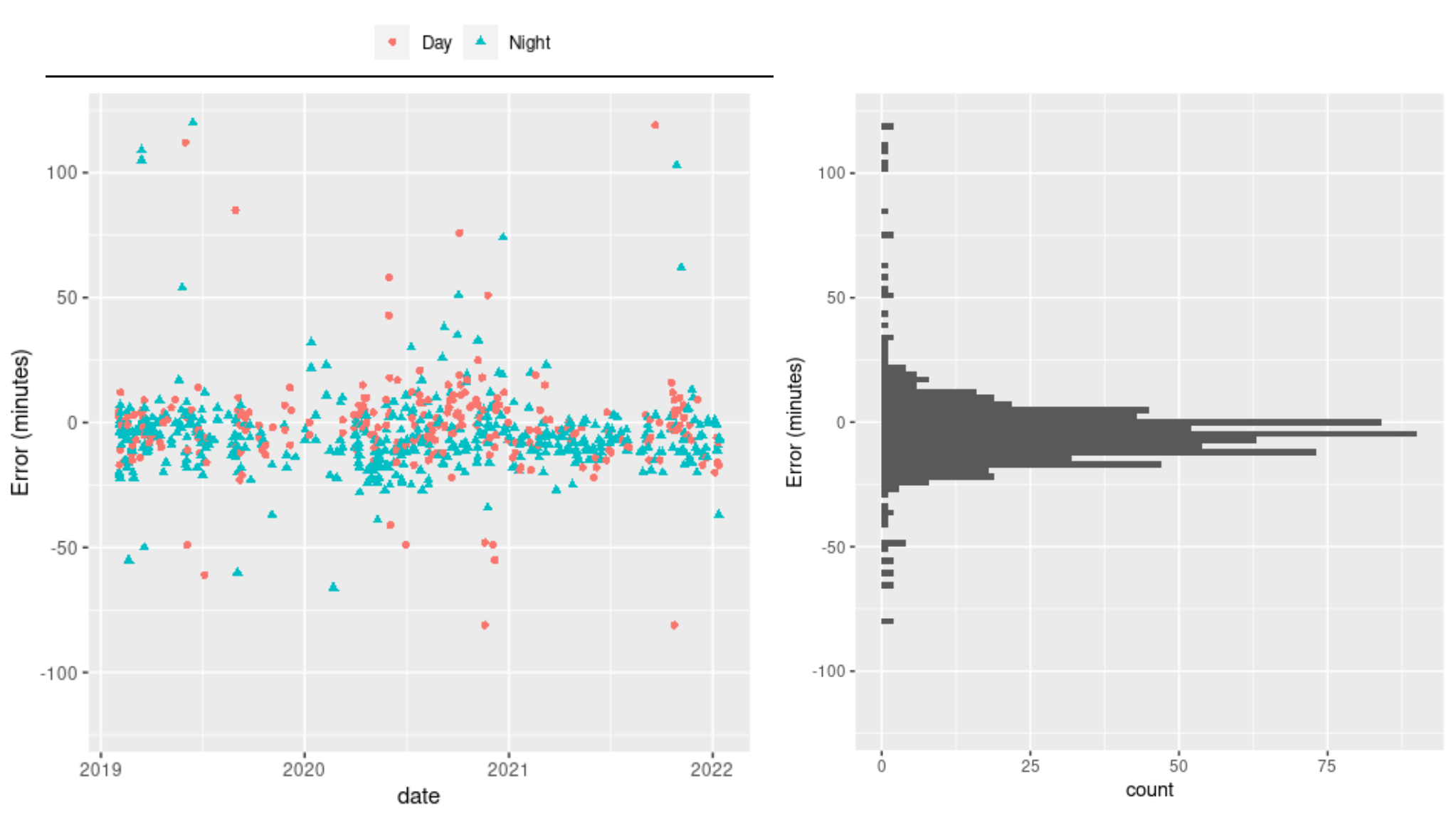}
    \caption{Difference between local mean solar time measured with the astrolabe and Eastern standard time over the course of the data collection (left) and histogram (right).}
    \label{fig:time_error}
  \end{center}
\end{figure}

The local mean solar time can be compared with my timezone time (Eastern standard time).
Using either the Sun or visible stars results in $n=765$ distinct time estimates (the remaining $773-765=7$ observing sessions are instances where not enough data were collected to uniquely determine the time).  For these estimates, the mean difference was  $-7.4$ minutes and the standard deviation was $57$ minutes.  A histogram of the differences is shown in Figure \ref{fig:time_error}, and is apparently close to normal.

The length of a daytime observing session is quite short (typically one or two elevation sightings, as indicated in Figure \ref{fig:dataset_summary}) and the length of a nighttime observing session is typically longer.
It would not be surprising if there were a difference in the time error for day versus night observing sessions.  Moreover, there is a significant difference between elevation errors for the Sun versus the stars (see Section \ref{sec:elevation_error}).  However, in the face of these potential problems, an ANOVA test shows no significant difference between time estimates derived from the Sun in the daytime versus the stars at night.

\begin{figure}
  \begin{center}
    \includegraphics[width=3.5in]{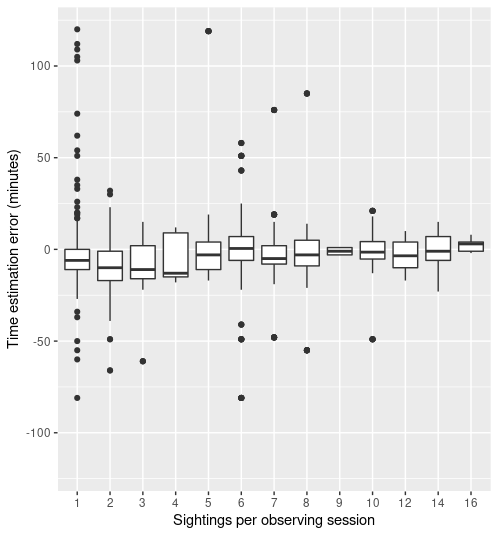}
    \caption{Difference between local mean solar time and Eastern standard time versus the number of sightings taken.}
    \label{fig:time_error_vs_sightings}
  \end{center}
\end{figure}

Each observing session lasted about 3--5 minutes.
For consistency, I recorded a single time from my computer's clock at the beginning of each observing session,
regardless of the overall length of the session.
However, when multiple celestial bodies are visible (usually at night), the time accuracy may be improved.
One visually splits the elevation errors across each of the visible stars on the front of the astrolabe \cite[II.5]{Chaucer_astrolabe}.
Both of these effects are present in the differences as a function of sighting count as shown in Figure \ref{fig:time_error_vs_sightings},
which does not indicate any strong trends.


\subsection{Longitude accuracy}

My computer's clock provides an accurate common time reference (Eastern Standard time),
which is aligned to the local mean solar time at $75^\circ$ West longitude.
The mean time difference I computed in Section \ref{sec:mean_time} can be used for the purpose of determining my longitude (as was described in Section \ref{sec:longitude}).
The difference of $-7.4$ minutes noted in Section \ref{sec:mean_time} translates to a longitude of $76.85^\circ$ West, which is therefore an estimate of my home longitude.
This estimate is about $13$ kilometers off from the true value.  That is quite good given the limited equipment I used!

\subsection{Right ascension errors}
\label{sec:ra_error}

One uses the timeseries of right ascension measurements of a planet to estimate its orbit using Kepler's laws,
as will be described in Section \ref{sec:orbits}.
Therefore, it is important to measure the accuracy of the right ascension measurements obtained with the astrolabe.

In addition to the right ascension measurements of each planet I obtained with the astrolabe,
I also predicted the expected right ascensions with a computer algorithm.
The computer algorithm is based upon Kepler's elliptical orbits for the planets with modern values of the \emph{orbital elements} that define them \cite{trefil2012space}.
Each right ascension obtained using the astrolabe can be compared with these expected values, resulting in Table \ref{tab:ra_errors}.

\begin{table}
  \begin{center}
    \caption{Errors in right ascension observations (hours)}
    \label{tab:ra_errors}
    \begin{tabular}{|l|r|r|r|}
    \hline
    Planet & $n$ & Mean & Std. dev.\\
    \hline
    \hline
    Venus&47&0.222&0.580\\
    \hline
    Mars&72&0.219&0.584\\
    \hline
    Jupiter&111&-0.195&1.30\\
    \hline
    Saturn&90&-0.825&0.651\\
    \hline
    \hline
    Overall&320&-0.00896&0.927\\
    \hline
  \end{tabular}
  \end{center}
\end{table}

Table \ref{tab:ra_errors} shows that my typical right ascension measurement error has a standard deviation of $0.927$ hours.
As mentioned in Section \ref{sec:ecliptic}, I included a correction to the right ascension of the Sun on the star chart.
According to the equation of time (Section \ref{sec:equation_of_time}),
the true right ascension can differ from uniform motion by $16$ minutes of right ascension ($0.26$ hours of right ascension),
or about $1/100$ of Earth's rotation period.
This quantity is well below the standard deviations listed in Table \ref{tab:ra_errors},
so we conclude that my including of the correction of the equation of time on the right ascension of the Sun is not measurable using the astrolabe.

\begin{figure}
  \begin{center} 
    \includegraphics[width=4in]{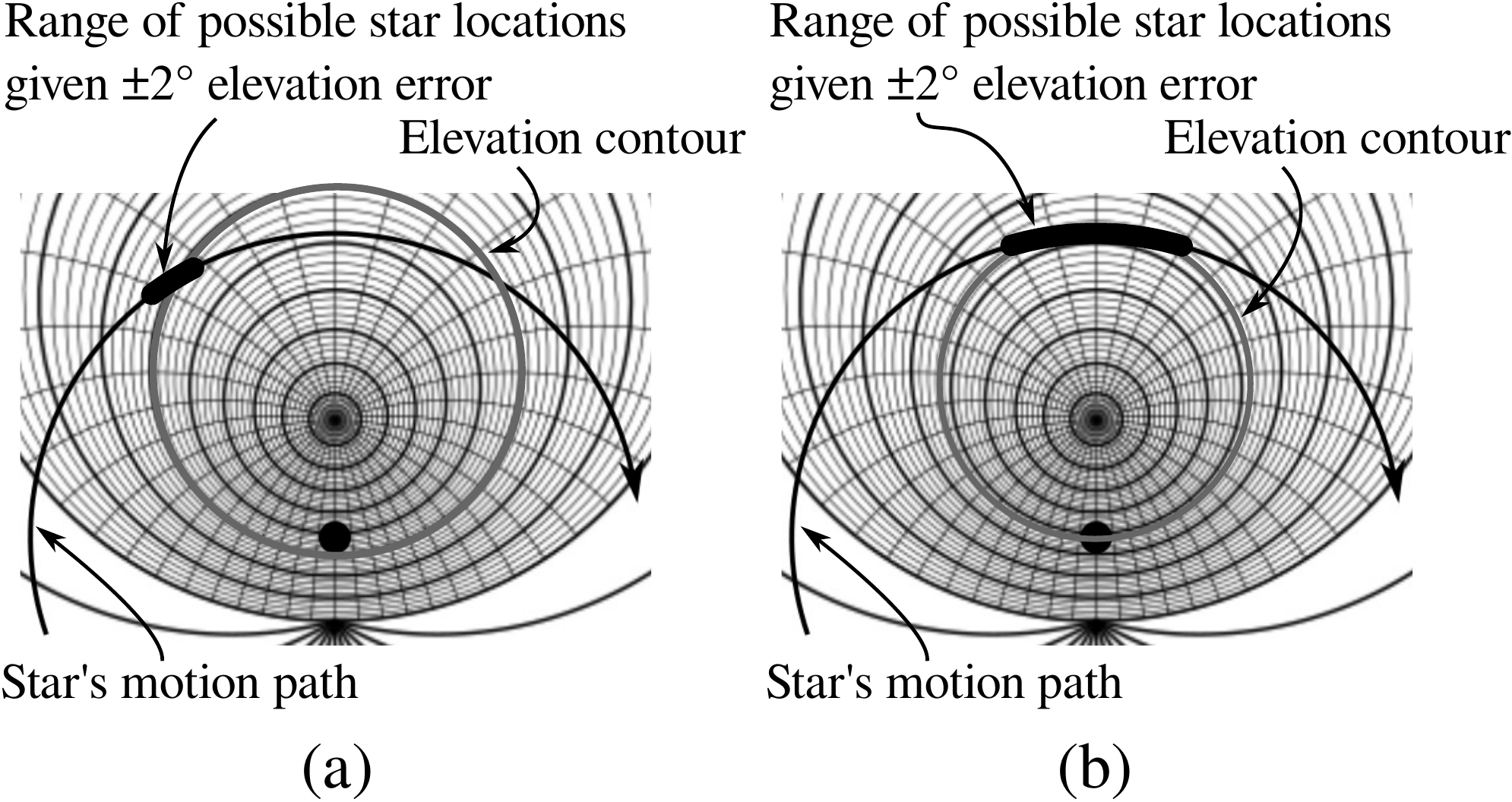}
    \caption{Star position uncertainty given a $\pm 2^\circ$ elevation error when (a) the star is far from due South or (b) near due South.}
    \label{fig:tangencies}
  \end{center}
\end{figure}

Chaucer claims in the last paragraph of \cite[II.3]{Chaucer_astrolabe} that right ascension errors can be noticeably worse when an object is nearly due South, because the elevation contours drawn on the astrolabe become tangent to the object's path of motion through the sky.  For instance, in Figure \ref{fig:tangencies}, the impact of a $\pm 2^\circ$ elevation error appears to depend significantly on whether the star is near due South (the vertical line of symmetry of the astrolabe) or not.  Indeed, Chaucer cautions the reader,

\begin{quote}
  \emph{But natheles, in general, wolde I warne thee for evere, ne mak thee nevere bold to have take a iust ascendent by thyn Astrolabie, or elles to have set iustly a clokke, whan any celestial body by which that thou wenest governe thilke thinges ben ney the south lyne.}
  
  \bigskip
  \noindent In general, I warn you not to measure a right ascension of a celestial body using your astrolabe---or use it to set your clock---if it is near due South.
\end{quote}

\begin{figure}
  \begin{center}
    \includegraphics[width=3.5in]{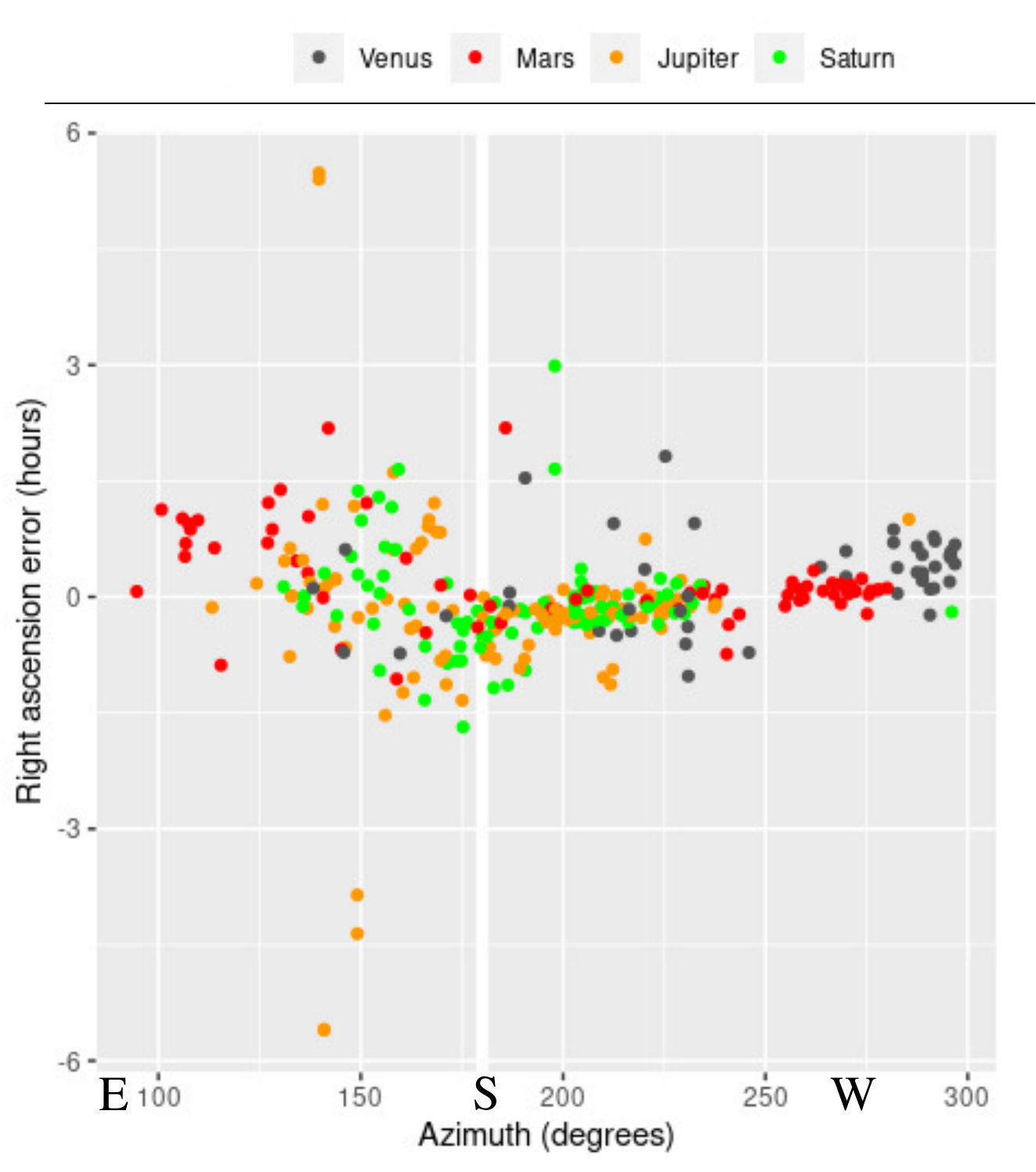}
    \caption{Right ascension error as a function of azimuth.  Due South ($180^\circ$) is marked with a vertical rule.}
    \label{fig:chaucer_hypothesis}
  \end{center}
\end{figure}

This claim can be tested using the data I have collected.  Figure \ref{fig:chaucer_hypothesis} displays right ascension error as a function of azimuth for the planets.  While there is a noticeable discontinuity at due South, the error does not change much as due South is approached from either side.  While the largest errors---which may be considered outliers---do seem more pronounced near due South, I think that perhaps Chaucer was being overly cautious!

The design of Chaucer's instrument was quite similar to mine, though he was based in Oxford, England.  Oxford is substantially further north than my home latitude, which does make the tangency between elevation and declination contours more severe.  It may be that Chaucer's advice is sound in his location, especially when the instrument is in the hands of a beginner.

The discontinuity is worth further consideration.
I suspect it is a systematic error caused by my observing habits.  Notice that the negative right ascension errors on both sides of due South appear somewhat similar, while the positive right ascension errors are dramatically worse for azimuths less than $180^\circ$.

Although I cannot say for certain, I believe an explanation for the discontinuity in Figure \ref{fig:chaucer_hypothesis} at $180^\circ$ could be due to buildings that blocked my view of the sky.
The walls of my house are closely aligned with North/South and East/West, so it is very easy to tell if an object has passed due South by simply sighting along one of the outer walls of my house.  If an object appears East of due South (so its azimuth is less than $180^\circ$), I am apparently less likely to check if this is truly the case.  Such an error (mistakenly choosing a right ascension that is currently East as opposed to West) would cause its right ascension error to be positive.

\section{Estimating the orbits of the planets}
\label{sec:orbits}
\begin{figure}
  \begin{center}
    \includegraphics[width=3.5in]{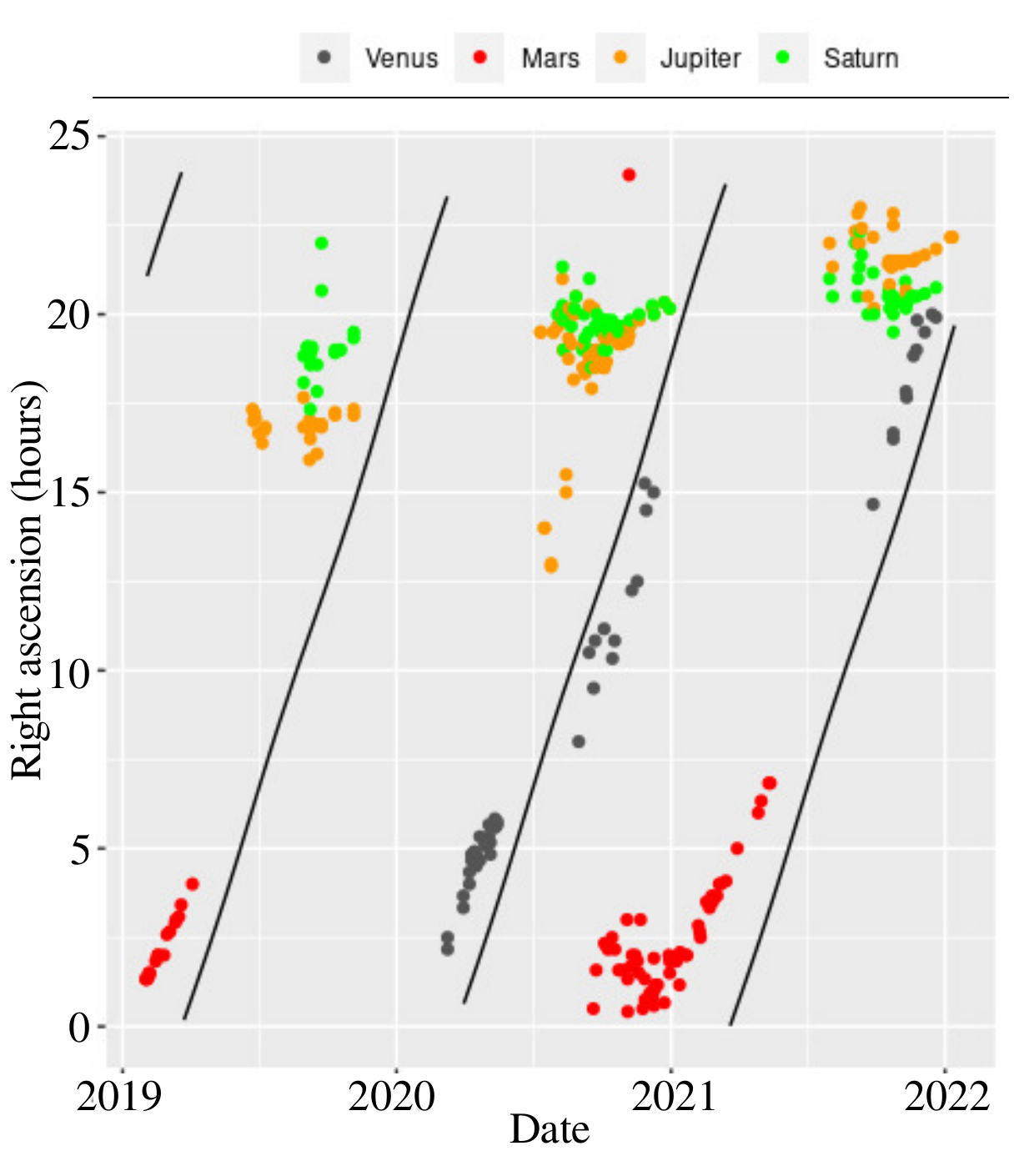}
    \caption{Observed right ascensions of the planets (colored points) and the Sun (black lines).}
    \label{fig:planetary_data}
  \end{center}
\end{figure}

From my location, I was able to observe Venus, Mars, Jupiter, and Saturn (see Figure \ref{fig:planetary_data} and Table \ref{tab:ra_errors}).  Although Mercury is certainly visible without optical aid, its maximum elevation is too low to be consistently observed from my location.  It is below the treeline or neighboring buildings.
I made no observations of Uranus or Neptune; both are difficult targets without a telescope!  

The data one can collect on the planets with an astrolabe are elevation sightings, which can be converted to right ascensions as explained in Section \ref{sec:predicting_true_locations}.  These are not sufficient to place the planets in their true locations in space, nor can they be used to determine their orbits without further information.  The key insight that allows one to make the leap from right ascensions to orbital dimensions is provided by Kepler's laws.   These laws connect orbital dimensions to orbital periods, and orbital periods can be derived from timeseries of right ascensions as will be explained below.

Kepler's third law states, ``The square of the orbital period of a planet is directly proportional to the cube of the semi-major axis of its orbit.''  In other words, if $r$ is the semi-major axis of the orbit and $T$ is its orbital period, 
\begin{equation}
  \label{eq:circular_radius}
  r^3 = k T^2.
\end{equation}
The derivation of Equation \eqref{eq:circular_radius} is quite well-known in the literature (for instance, see \cite{Mosna_2014} for a very elementary derivation).

\begin{figure}
  \begin{center}
    \includegraphics[width=2.5in]{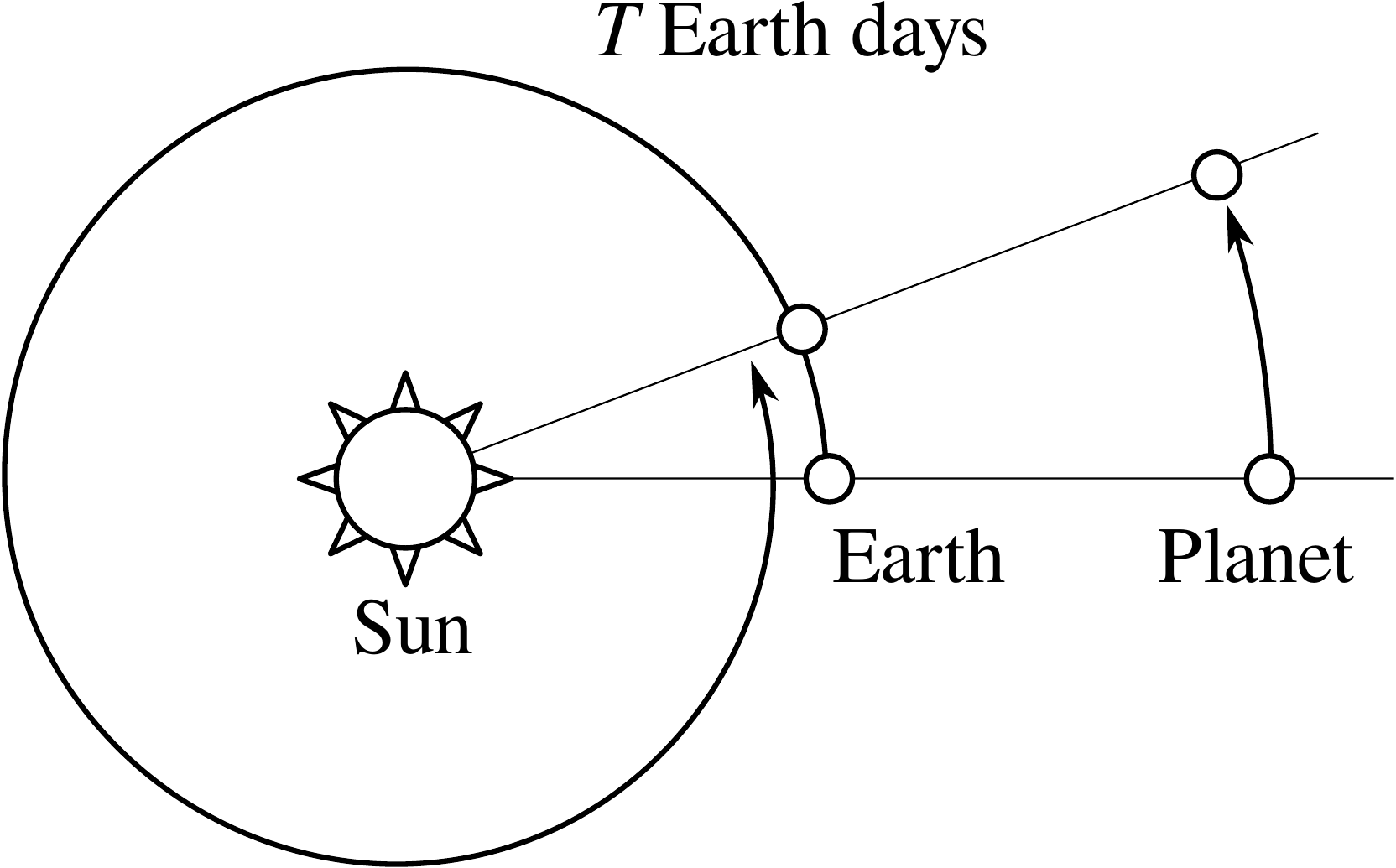}
    \caption{The synodic period for a planet whose orbit is outside Earth's orbit is the time between solar oppositions.}
    \label{fig:synodic}
  \end{center}
\end{figure}

Directly measuring the orbital period of a planet is complicated by the fact that both the planet and the Earth are in motion.
Fortunately, the planet's orbital period can be estimated using a geometric trick that works because the planets lie close to the ecliptic.
There are distinct times when the Sun, the Earth, and the planet are (nearly) colinear in space.
These times are called \emph{solar oppositions} or \emph{solar conjunctions},
depending on whether the planet and the Sun are on opposite or the same side of the Earth, respectively.
The path of a planet and the Earth between two consecutive solar oppositions is shown in Figure \ref{fig:synodic}.
The time interval between two consecutive conjunctions or oppositions is called the \emph{synodic period} of the planet.

Kepler's second law implies that if a planet's orbit is circular,
then the planet moves at a constant angular speed over its entire orbit.
All four planets I observed have orbits that are close enough to circular that we can safely assume that their angular speeds are constant.

Over one synodic period $T$ (days), the angular distance travelled by the Earth is approximately
\begin{equation*}
  2 \pi \frac{T}{365} \text{ radians}.
\end{equation*}
Over the same period of time, a planet on a circular orbit outside Earth's orbit will travel significantly slower,
and has only travelled
\begin{equation*}
  2 \pi \frac{T-365}{365} \text{ radians}.
\end{equation*}
Let us use the Earth's orbit as a yardstick for both radius and period.
That is, for the Earth, $r=1$ astronomical unit (AU) and $T=365$ days, so that
\begin{equation*}
  k=(1/365) (\text{AU}^3/\text{day}^2).
\end{equation*}
Therefore, Kepler's third law asserts that for an \emph{outer planet}---one whose orbit lies outside Earth's orbit---the semimajor axis $r$ in astronomical units is given by
\begin{equation}
  \label{eq:outer_radius}
  r = \left(\frac{T}{T-365}\right)^{2/3} \text{ AU}.
\end{equation}

The situation is reversed for an \emph{inner planet}, one whose orbit lies entirely within Earth's orbit.  In this case, the planet never exhibits solar oppositions, since the planet cannot be on the opposite side from the Earth as the Sun.  As a result, the planet oscillates between apparent solar conjunctions on opposite sides of the Sun.  The synodic period $T$ is the time in Earth days between two of the same kind of conjunction.  Again, during this time, the Earth has traveled approximately
\begin{equation*}
  2 \pi \frac{T}{365} \text{ radians}.
\end{equation*}
However, the planet has traveled
\begin{equation*}
  2 \pi \frac{T+365}{365} \text{ radians}.
\end{equation*}
Therefore, for an inner planet, its semimajor axis $r$ is given by the formula
\begin{equation}
  \label{eq:inner_radius}
  r = \left(\frac{T}{T+365}\right)^{2/3} \text{ AU}.
\end{equation}

\subsection{Using the synodic period}

The synodic period of a planet is the key to unlocking its orbital dimensions and can be derived from timeseries of right ascensions.  Because the synodic period relates the Earth's and the planet's orbits to each other using conjunctions or oppositions, it helps to subtract the planet's right ascensions from the Sun's right ascensions.  The resulting \emph{Sun-relative right ascensions} have the useful property that $0$ hours correspond to conjunctions, while $\pm 12$ hours correspond to oppositions.

\begin{figure}
  \begin{center}
    \includegraphics[width=5in]{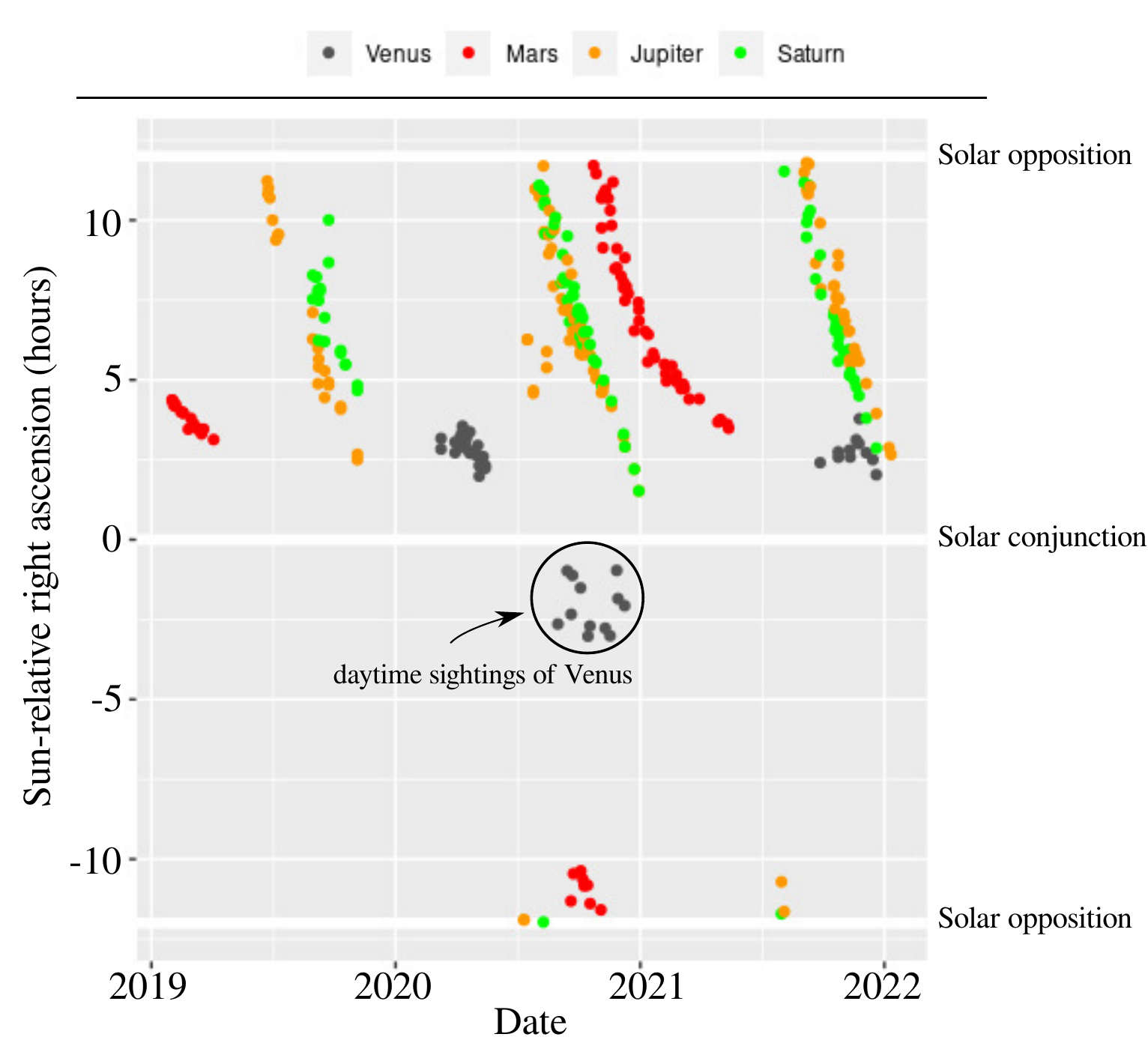}
    \caption{Sun-relative right ascensions of the planets.}
    \label{fig:sun_relative_data}
  \end{center}
\end{figure}

Figure \ref{fig:sun_relative_data} shows the same data as Figure \ref{fig:planetary_data}, but as Sun-relative right ascensions.  The representation in Figure \ref{fig:sun_relative_data} reveals solar conjunctions and oppositions more clearly, and makes it much easier to estimate synodic periods.

One thing that is immediately visible in Figure \ref{fig:sun_relative_data} is that Venus \emph{never appears in solar opposition} in my data.
Especially taking into account the daytime sightings, it appears on both sides of the Sun.
Therefore, one would have expected that \emph{if they were possible, solar oppositions of Venus would be present}.
We can also conclude that Venus is an inner planet, because it is never appeared ``outside'' Earth's orbit.
Moreover, a hint of sinusoidal variation is visible for the Venus sightings in Figure \ref{fig:sun_relative_data}.
From this, it seems reasonable to claim that \emph{I have directly observed Venus orbiting the Sun instead of the Earth}.
Unassailable proof of this fact is credited to Galileo's telescopic observation of the phases of Venus in time with this sinusoial variation.
This does not, by the way, establish that the Sun is at the center of the solar system, only that Venus orbits the Sun!
Because the other three planets exhibit solar oppositions, we must conclude that they are outer planets.

The synodic period is measured between consecutive oppositions or conjunctions,
though it is clear from Figure \ref{fig:sun_relative_data} that I did not observe many of these events directly.
Fortunately, the synodic period can also be estimated by considering the time interval between recurrence of a Sun-relative right ascension (that is, directly from what is shown in Figure \ref{fig:sun_relative_data}).

As an example, I saw Saturn on 6 October 2020 at Sun-relative right ascension $6.1$ hours, and again at that same Sun-relative right ascension on 23 October 2021.  Both of these observations correspond to points in Figure \ref{fig:sun_relative_data}.  Together, these two observations suggest a synodic period estimate of $382$ days.  Using this as the synodic period results in an estimate of $7.96$ AU for Saturn's semi-major axis.  Over the entirety of my data, there are many such possible recurrences.

\begin{table}
  \begin{center}
    \caption{Orbital semi-major axes of the planets (AU)}
    \label{tab:planet_radii}
    \begin{tabular}{|l|r||r|r|r|r|}
    \hline
    Planet &True value& $n$ & Mean obs. &Median obs.& Std. dev.\\
    \hline
    \hline
    Venus&0.723&136&0.718 &0.720 &0.00992 \\
    \hline
    Mars&1.52&101&1.57 &1.57 &0.0277 \\
    \hline
    Jupiter&5.20&653&5.25 &4.87 &1.88 \\
    \hline
    Saturn&9.55&440&8.46 &7.66 & 3.24\\
    \hline
  \end{tabular}
  \end{center}
\end{table}

\begin{figure}
  \begin{center}
    \includegraphics[width=5in]{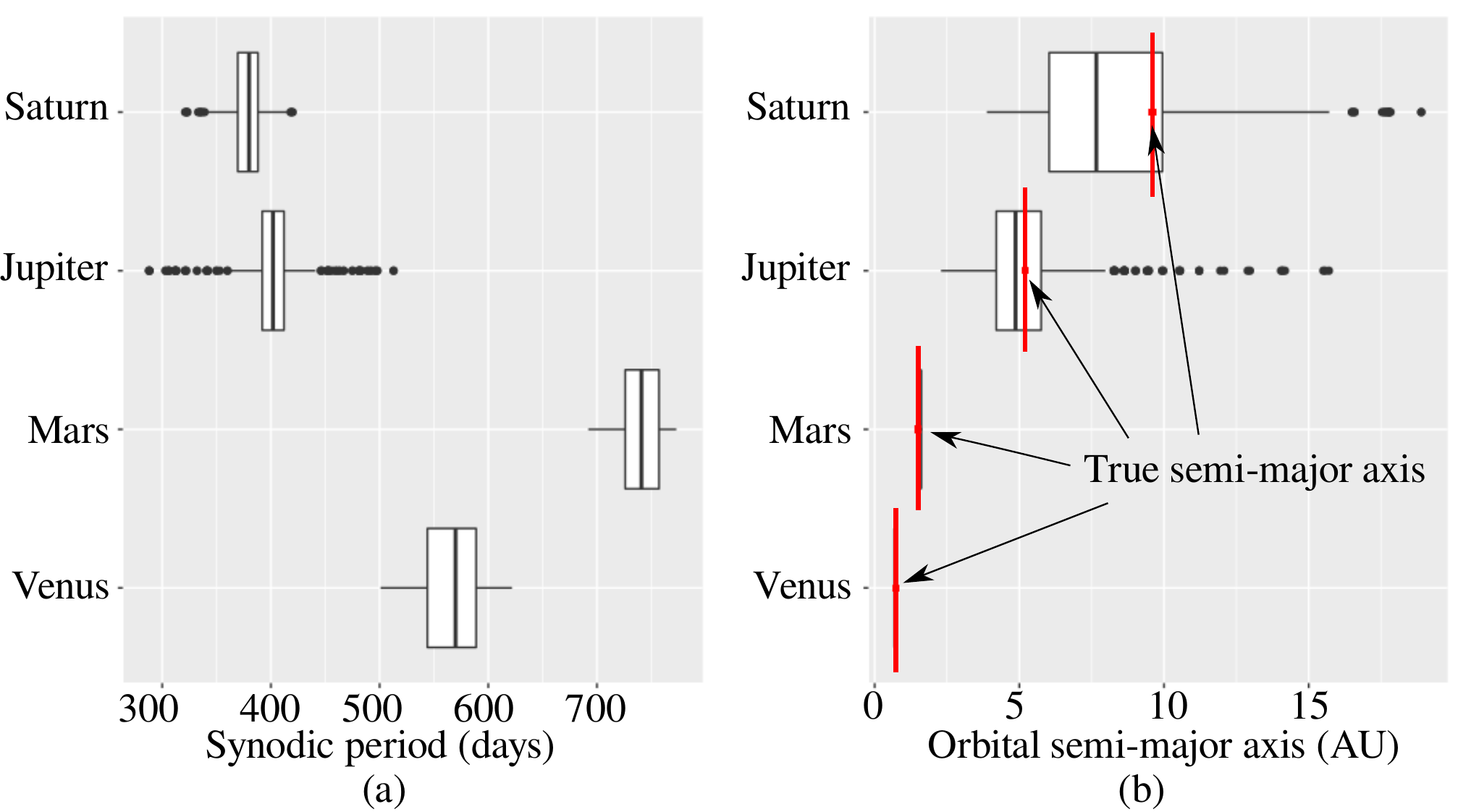}
    \caption{Distribution of estimate for (a) the synodic periods and (b) the orbital semi-major axes of the four planets observed in the dataset.  Accepted figures for the orbital semi-major axes are as follows: Venus $0.72$ AU, Mars $1.5$ AU, Jupiter $5.2$ AU, Saturn $9.6$ AU.  Notice that the standard deviations of semi-major axis for Venus and Mars are quite small!}
    \label{fig:planetary_summary}
  \end{center}
\end{figure}

Although a given Sun-relative right ascension may not recur exactly,
there are many cases where a similar Sun-relative right ascension occurs later in the data.
To capture these ``approximate recurrences,'' Sun-relative right ascensions were grouped into $2$ hour bins.
Each pair of observations in each bin corresponds to a possible ``approximate recurrence'' of that Sun-relative right ascension,
and the time interval between them is an estimate of the planet's synodic period.
Figure \ref{fig:planetary_summary}(a) shows the distribution of these estimates for the synodic periods.

Because binning also collects observations of the same opposition or conjunction, rather than a recurrence, short intervals must be removed.
Intervals greater than $200$ days between observations in these bins correspond to instances where the planet had re-appeared in the same relative location and were used.  All of these time intervals between pairs of observations in these bins were aggregated into a distribution, shown in Figure \ref{fig:planetary_summary}(a).  Each of these time intervals was used to estimate the orbital semi-major axis using Equation \eqref{eq:outer_radius} (for the outer planets) or \eqref{eq:inner_radius} (for Venus).  The resulting estimates are shown in Figure \ref{fig:planetary_summary}(b) and Table \ref{tab:planet_radii}.  Note that the counts in the $n$ column of Table \ref{tab:planet_radii} correspond to \emph{pairs} of sightings, and so differ from the number of sightings of each planet in Table \ref{tab:ra_errors}.

\subsection{The case of Venus}

\begin{figure}
  \begin{center}
    \includegraphics[width=5in]{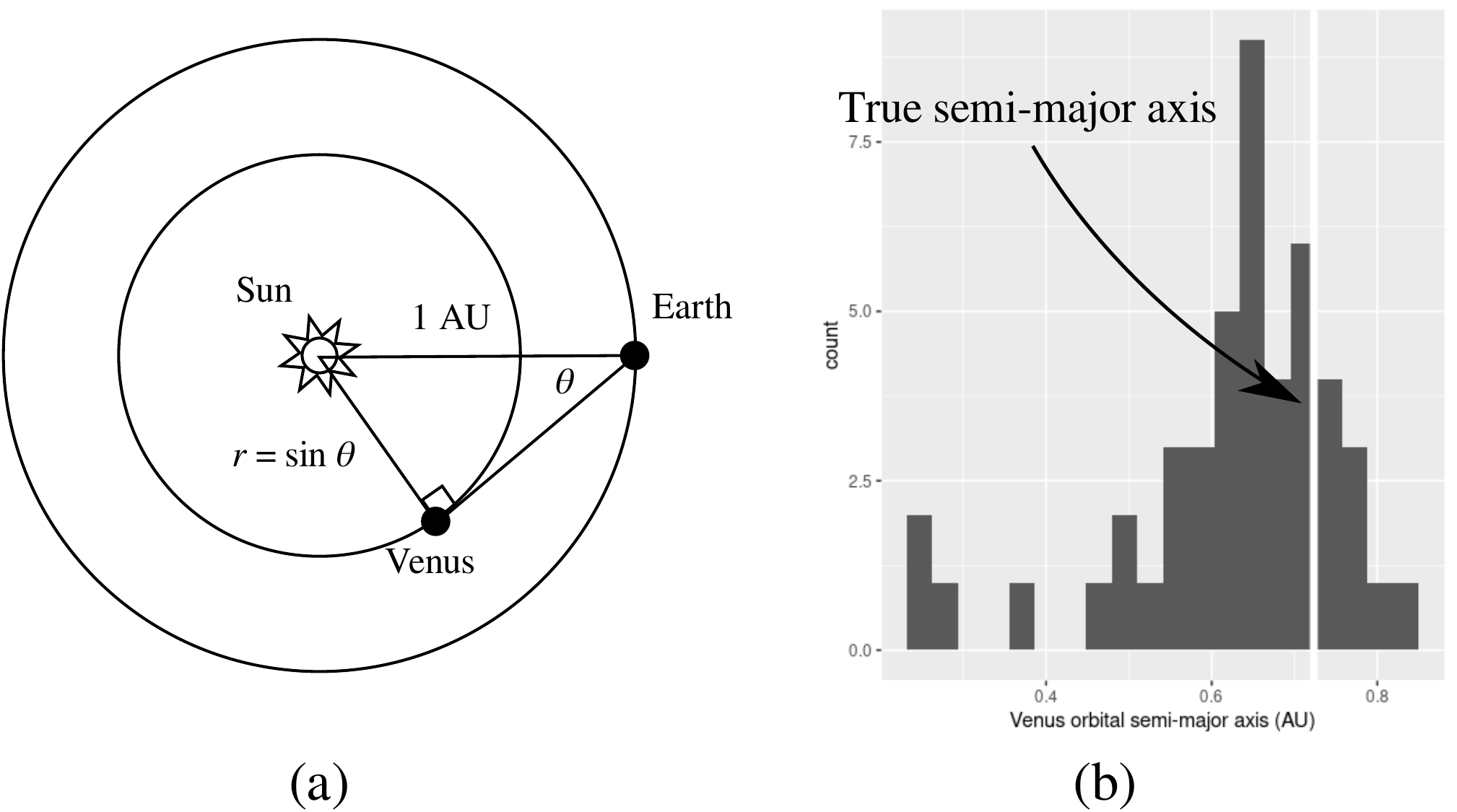}
    \caption{(a) The geometry of Venus's and Earth's orbits, where $\theta$ is the observed Sun-relative right ascension of Venus (see Figure \ref{fig:sun_relative_data}), and (b) Estimates of Venus's orbital semi-major axis obtained from the data ($n=47$) shown in Figure \ref{fig:sun_relative_data}.  The mode is $0.650$ AU, and the true semi-major axis of $0.723$ AU is indicated.}
    \label{fig:venus_data}
  \end{center}
\end{figure}

Because I collected numerous observations of Venus on both sides of the Sun, we are able to crosscheck our estimates of its orbit by another method that does not rely upon Kepler's law.  We can use geometry to directly estimate the orbital radius from Venus's largest Sun-relative right ascension, if we assume that its orbit is circular.  The geometry of this situation is shown in Figure \ref{fig:venus_data}(a), where $\theta$ is the maximum Sun-relative right ascension.  We can read off the maximum values of Sun-relative right ascension directly from Figure \ref{fig:sun_relative_data}.  

The geometry in Figure \ref{fig:venus_data}(a) indicates that we only need to estimate the \emph{maximum} Sun-relative right ascension.
We can simply use \emph{every} observation of Venus to derive an estimate of Venus's orbital radius by taking the mode of the resulting distribution.
If there were no errors, the distribution of these observations would have a single sharp peak at the maximum Sun-relative right ascension, with no observations larger than this value.  

Figure \ref{fig:venus_data}(b) shows the distribution of estimates obtained by computing the sine of the Sun-relative right ascension of Venus, in which a total of $n=47$ observations were available.  The distribution of these radius estimates has a fairly heavy left tail, corresponding to the points in time where Venus was not at its maximum apparent distance from the Sun.  This tail is not very relevant, since we are most interested in the \emph{maximum} value represented.  To estimate maximum, we can use the mode, which is $0.650$ AU.  The standard deviation for the observations is $0.131$ AU. The true value of $0.723$ AU is within one standard deviation of $0.650$ AU, so we conclude that using geometry alone yields a good estimate of Venus's orbital radius.  

\subsection{Scale model of the solar system}
\label{sec:orrery}

The astrolabe data can also be used to create a model of the solar system that places the planets in their true locations with respect to the Sun.

To place a planet at its location based upon its right ascension,
one needs to calculate the intersection between the planet's orbit with a ray of constant right ascension that starts at the earth.
 
\begin{figure}
  \begin{center}
    \includegraphics[width=2.5in]{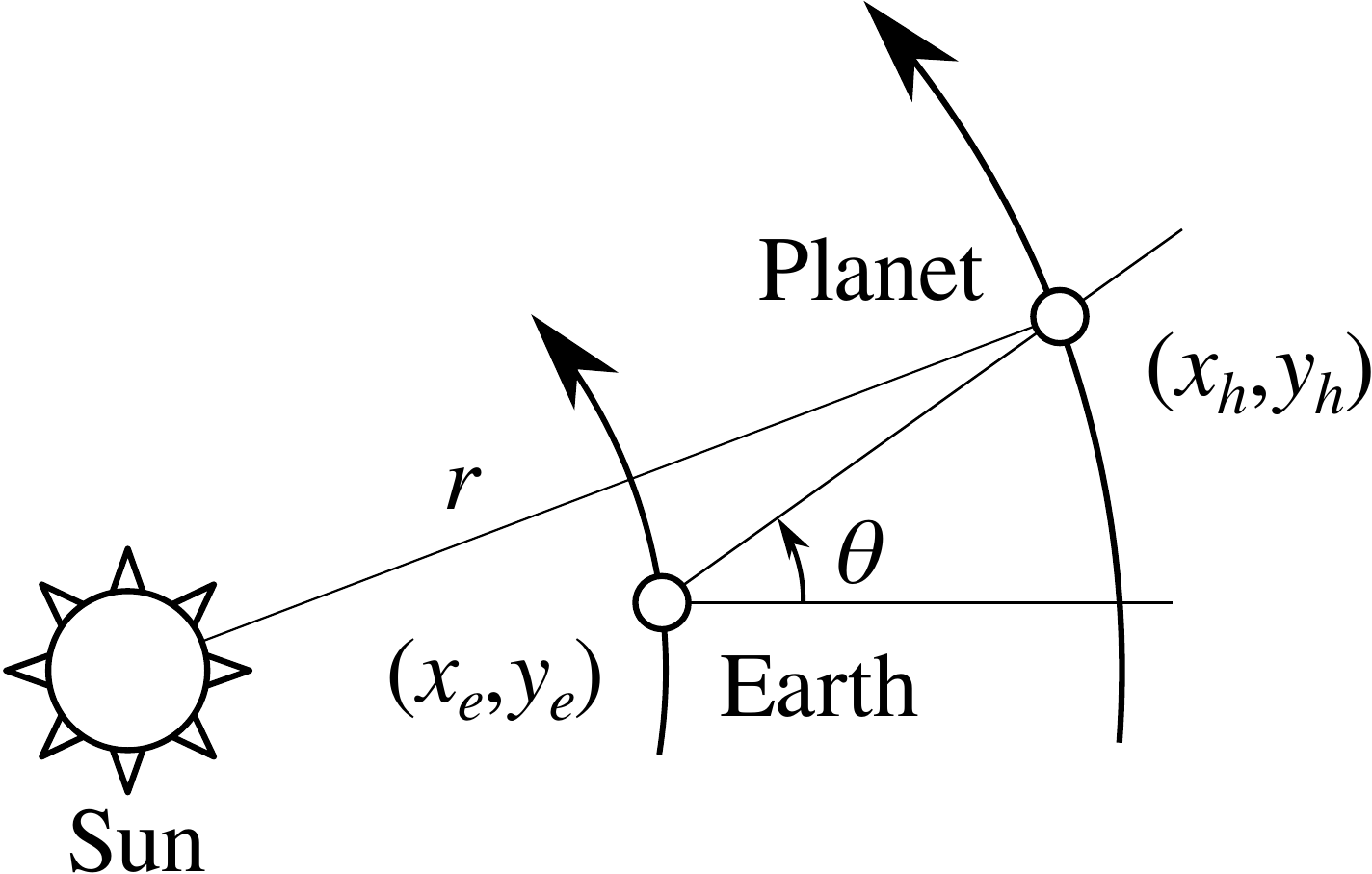}
    \caption{Relating the position of a planet to the Earth.}
    \label{fig:planet_position}
  \end{center}
\end{figure}

Suppose the Earth is located at $(x_e,y_e)$, the planet is located at $(x_h,y_h)$, and the Sun is located at the origin.  If the observed right ascension is $\theta$, as shown in Figure \ref{fig:planet_position}, then the location of the planet is at
\begin{equation}
  \label{eq:p_eqn}
    x_h = x_e + t \cos\theta, \; y_h = y_e + t \sin\theta
\end{equation}
for some $t>0$, which must be determined.  If the planet's orbit is circular with radius $r$, this means that
\begin{equation*}
  \begin{aligned}
    r^2 &= x_h^2 + y_h^2\\
    &= (x_e + t \cos\theta)^2 + (y_e + t \sin\theta)^2\\
    &= x_e^2 + 2 tx_e \cos\theta + t^2 \cos^2\theta + y_e^2 + 2 y_e t \sin\theta + t^2\sin^2\theta\\
    &= t^2 + 2 t (x_e \cos\theta + y_e \sin\theta) + x_e^2 +y_e^2 \\
    0 &= t^2 + 2 t (x_e \cos\theta + y_e \sin\theta) + 1-r^2.
  \end{aligned}
\end{equation*}
Solving for $t$ yields
\begin{equation}
  \label{eq:t_eqn}
  t = -(x_e \cos\theta + y_e \sin\theta) \pm \sqrt{(x_e \cos\theta + y_e \sin\theta)^2-(1-r^2)}.
\end{equation}
There are two possible solutions for $t$, corresponding to two intersections of the ray of constant right ascension with the planet's orbit.
Since the Earth's orbit is entirely contained within the orbit of an outer planet, these two intersections will be on opposite sides of the Earth.
As a result, only the positive root needs to be taken for outer planets, since the negative root will result in $t<0$.
For inner planets, there is an unresolvable ambiguity given the data provided, since both intersections are consistent with the right ascension given.

\begin{figure}
  \begin{center}
    \includegraphics[width=3.5in]{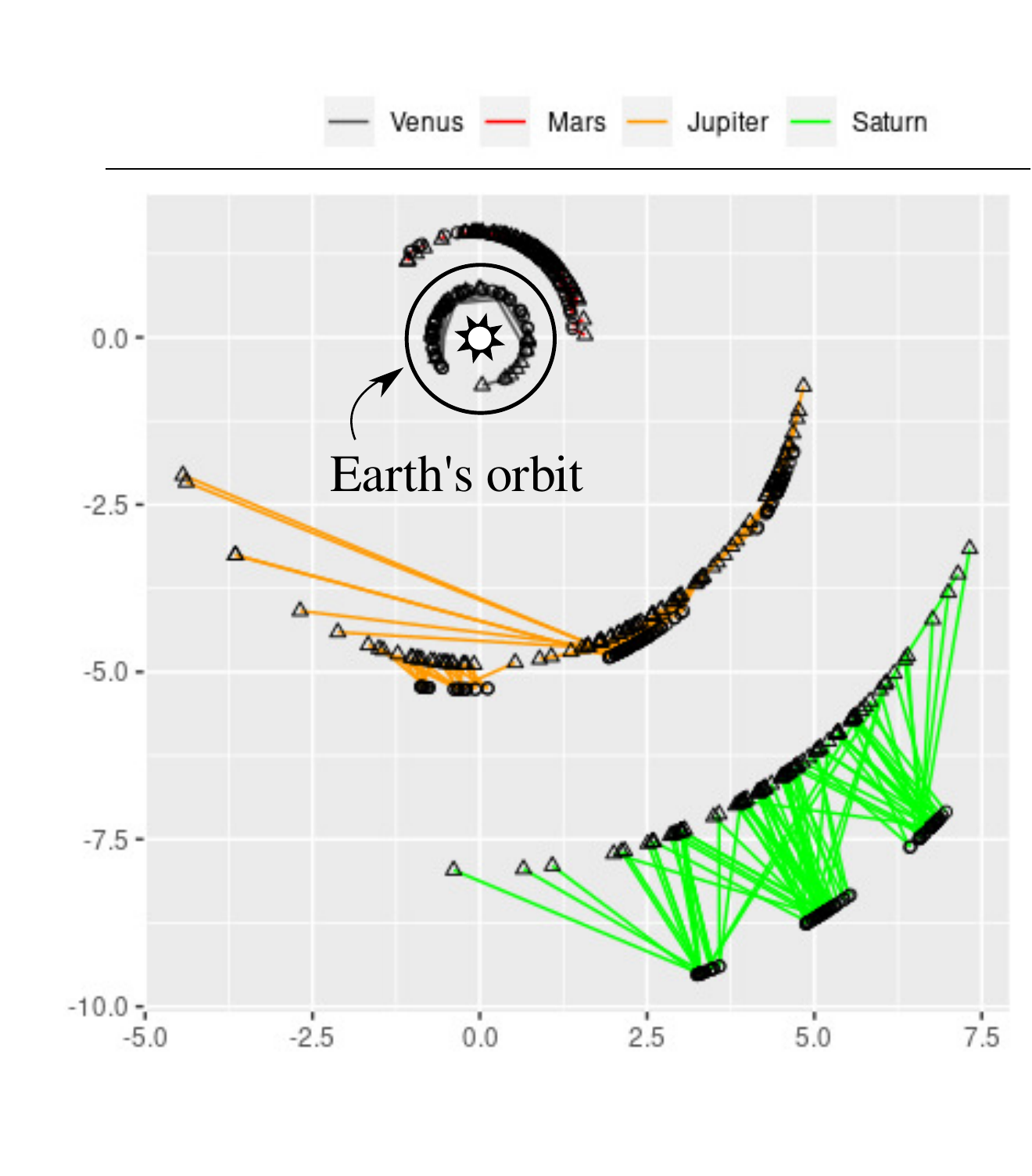}
    \caption{To-scale model of the solar system showing each planetary obseration.  True planetary positions are shown with circles, which are connected to the corresponding observations (triangles) with line segments.  Axes are in AU.}
    \label{fig:orrery}
  \end{center}
\end{figure}

Figure \ref{fig:orrery} shows the positions of the planets in the solar system versus their observed positions.  The radii for the observed positions are derived using Equations \eqref{eq:p_eqn} and \eqref{eq:t_eqn}, using the median observed planet orbital semi-major axis in Table \ref{tab:planet_radii} as the radius of the planet's circular orbit.  Observations of Jupiter and Saturn are considerably spread from their true locations.  
For instance, the standard deviation of right ascension errors for Saturn is $0.65$ hours = $9.75^\circ$.  That means that the 95\% confidence interval for Saturn right ascensions is nearly $\pm 20^\circ$ from the true value. This is largely in agreement with the widest outliers shown Figure \ref{fig:orrery}.

\section{Conclusion}

Table \ref{tab:planet_radii} affirms that given enough data (a few years' worth), an astrolabe can help you measure the solar system, even if you don't have dark skies!  I have demonstrated that it is possible to measure:
\begin{enumerate}
\item one's latitude and longitude (to around $1^\circ$, if you are careful and patient),
\item the time to within $30$ minutes consistently, often much better,
\item the current locations of objects in the sky, and
\item the size of the orbits of the nearby planets.
\end{enumerate}

You can even suggest (based on the observations of Venus) that some planets orbit the Sun, and not the Earth!
When supplemented with daytime observations,
the Sun-relative right ascensions of Venus indicate that no oppositions ever occur, a clear signal that it orbits the Sun.
Admittedly, to make the daytime observations of Venus I had to use binoculars to ``acquire'' the sighting, even though the astrolabe was used to measure the elevation.  This is perhaps an overly modern advantage!

Although I highly recommend constructing and using an astrolabe, even without one, the reader should go outside at night and look at the sky anyway.
There are many worthy things to be learned!
Although I am not done observing, the data I collected can be used to answer many other questions beyond those described here.
For instance, I also collected sightings of the Moon, including its phase.
Can these determine the Moon's orbit closely enough to be able to predict its location in the future?
How does the error depend on the forecast length?

There is also the possibility of using a telescope in conjunction with the astrolabe to collect measurements of fainter objects.
Using a telescope, I have already collected a few observations of the phases of Venus.
Even though they cannot be seen by the unaided eye, even a small telescope allows one to track the planets Uranus and Neptune.
I have seen Uranus through a telescope a handful of times and perhaps even Neptune, but not enough yet to do any systematic analysis.

If one is especially ambitious,
the techniques explained in this article could be made to work for minor planets, such as Vesta, Pluto, or Ceres.
Some of the minor planets are sufficiently close to the ecliptic that the techniques here can work without much change.
For minor planets whose orbital planes are not as closely aligned with the ecliptic as the major planets,
one could use Chaucer's instructions to determine their declinations \cite[II.30]{Chaucer_astrolabe}.
With these in hand, Kepler's laws could help predict the path of minor planets as they move through the solar system.
This might be a challenging and rewarding exercise to attempt!

\section*{Dataset availability}

All of the data and analyses described in this article are freely available \cite{Robinson_astrolabe_analysis}.
The reader is also encouraged to make their own astrolabe.  Plans for the astrolabes used to collect the data in this article are freely available \cite{Robinson_large_astrolabe,Robinson_small_astrolabe}.

\section*{Acknowledgments}

The author would like to thank the directors of American University's Design and Build Lab, Kristof Aldenderfer and Gustavo Abbott, for their assistance in developing the laser-cut astrolabes used to collect the data described in this paper.

\bibliographystyle{plainnat}
\bibliography{astrolabe_bib}
\end{document}